%% file: 0.main.tex
\documentclass[10pt,journal]{IEEEtran}
\usepackage{algorithm}
\usepackage{hyperref}
\usepackage{multirow}
\usepackage{bbding}
\usepackage{color}
\usepackage{amsmath,amssymb,amsfonts}
\usepackage{amsthm}
\usepackage{graphicx}
\usepackage{setspace}
\usepackage{multicol}
\usepackage{multirow}
\usepackage{cite}
\usepackage{float}
\usepackage{color}
\usepackage{alltt, listings}
\usepackage{subfigure}
\usepackage{algpseudocode}
\usepackage{wrapfig}
\usepackage{url}
\usepackage{bm}

\newtheorem{assumption}{Assumption}
\newtheorem{theorem}{Theorem}
\newtheorem{lemma}{Lemma}

\newtheorem{remark}{Remark}
\theoremstyle{definition}
\newtheorem{definition}{Definition}

\newcommand{\nop}[1]{}
\def\BibTeX{{\rm B\kern-.05em{\sc i\kern-.025em b}\kern-.08em
    T\kern-.1667em\lower.7ex\hbox{E}\kern-.125emX}}

\begin {document}

\title {Layered Randomized Quantization for Communication-Efficient and Privacy-Preserving Distributed Learning}

\author{Guangfeng Yan*, Tan Li*, Tian Lan, Kui Wu and Linqi Song 

\thanks{*G. Yan and T. Li contributed to the work equally and should be regarded as co-first authors. (Corresponding author: Linqi~Song)}

\thanks{G. Yan, L. Song are with the Department of Computer Science, City University of Hong Kong, and City University of Hong Kong Shenzhen Research Institute, Shenzhen, China. (e-mail: gfyan2-c@my.cityu.edu.hk, linqi.song@cityu.edu.hk)} 
\thanks{T. Li is with the Department of Computer Science, The Hang Seng University of Hong Kong. (e-mail: tanli6-c@my.cityu.edu.hk)}
\thanks{T. Lan is with the Department of Electrical and Computer Engineering, George Washington University. (e-mail: tlan@gwu.edu)}
\thanks{K. Wu is with the Department of Computer Science, University of Victoria. (e-mail: wkui@uvic.ca)}

}

\maketitle

\begin{abstract}
Next-generation wireless networks, such as edge intelligence and wireless distributed learning, face two critical challenges: communication efficiency and privacy protection. In this work, our focus is on addressing these issues in a distributed learning framework. We consider a new approach that simultaneously achieves communication efficiency and privacy protection by exploiting the privacy advantage offered by quantization.  Specifically, we use a quantization scheme called \textbf{Gau}ssian \textbf{L}ayered \textbf{R}andomized \textbf{Q}uantization (Gau-LRQ) that compresses the raw model gradients using a layer multishift coupler. By adjusting the parameters of Gau-LRQ, we shape the quantization error to follow the expected Gaussian distribution, thus ensuring client-level differential privacy (CLDP). We demonstrate the effectiveness of our proposed Gau-LRQ in the distributed stochastic gradient descent (SGD) framework and theoretically quantify the trade-offs between communication, privacy, and convergence performance. We further improve the convergence performance by enabling dynamic private budget and quantization bit allocation. We achieve this by using an optimization formula that minimizes convergence error subject to the privacy budget constraint. We evaluate our approach on multiple datasets, including MNIST, CIFAR-10, and CIFAR-100, and show that our proposed method outperforms the baselines in terms of learning performance under various privacy constraints. Moreover, we observe that dynamic privacy allocation yields additional accuracy improvements for the models compared to the fixed scheme. 
\end{abstract}

\begin{IEEEkeywords}
Distributed Learning, Communication Efficiency, Quantization, Privacy
\end{IEEEkeywords}

\input{Introduction}

\input{2.Related_Work}

\input{3.Problem_Formulation_new}
\input{4.LRQ-SGD_new}
\input{5.DLRQ-SGD}

\input{6.Discussion}
\input{7.Experiments}
\input{Conclusion}
\bibliographystyle{IEEEtran}
\bibliography{mybib}
\newpage
\input{Appendix}

\end{document}

%% file: Introduction.tex
\section{Introduction}
\label{introduction}


One of the key features of the next-generation wireless networks is the integration of communication networks and distributed systems with privacy protection. Wireless distributed learning allows the training process to be carried out across multiple machines or nodes. By leveraging computational resources from multiple machines, distributed learning can train larger models or process larger data sets in a shorter time frame~\cite{deng2020edge}. Distributed Stochastic Gradient Descent (Distributed SGD) is one of the commonly used algorithms \cite{}. Despite the apparent advantages, distributed learning is faced with three critical concerns, i) \textit{Communication efficiency}: Due to the sheer volume of devices involved, nodes within this learning setup require data sharing or parameter updates, and these processes may contribute to substantial communication overhead, potentially imposing inhibitive bottlenecks on the system~\cite{tang2020communication,tao2018esgd} ii) \textit{Privacy}: Exchanging model parameters or even just gradients may lead to the privacy leakage of local data~\cite{zhu2020deep,fredrikson2015model,nasr2019comprehensive}. iii) \textit{Convergence performance}: Unlike centralized training, distributed learning often requires extended time to achieve model convergence. Furthermore, methods aimed at improving communication efficiency or privacy, such as lossy compression and disrupting gradient, may result in additional training noise/errors that negatively impact convergence.

Indeed, a vast amount of research has been dedicated to addressing the aforementioned issues. For instance, compression techniques have been utilized to curtail communication overhead in wireless distributed learning~\cite{alistarh2017qsgd, basu2019qsparse, nadiradze2021asynchronous}. The issue of privacy leakage has been tackled by introducing noise to disrupt raw gradient updates~\cite{dwork2014algorithmic, bonawitz2017practical}. In addition, several studies~\cite{zong2021communication,mohammadi2021differential,agarwal2018cpsgd,agarwal2021skellam} have striven to tackle communication efficiency and privacy concerns simultaneously. 
However, they merely combine the techniques highlighted above. For instance, the authors in~\cite{zong2021communication,mohammadi2021differential} compress the Gaussian DP noise perturbed gradient without providing a theoretical guarantee for the final model's performance. 
Different from them, the authors in~\cite{agarwal2018cpsgd,agarwal2021skellam} employ various forms of discrete additive noise in conjunction with quantized gradient. However, these methods do not fully exploit the privacy advantages offered by compression itself. This simple amalgamation of disparate techniques inherently encounters issues, such as high computational overhead due to the additional processing requirements. 
Consequently, an intriguing question remains unanswered: \textit{\textbf{Can compression techniques alone concurrently ensure communication efficiency, privacy, and guaranteed model convergence performance?}}

Some existing works have tried to answer this problem but with limited success. Specifically, the works ~\cite{amiri2021compressive,li2022communication} simultaneously achieve communication efficiency and privacy for distributed learning by leveraging quantization. However, their methods only showed empirical benefits. The most related works to us are~\cite{yan2023killing, youn2023randomized}. The authors studied the privacy properties of compression and showed that the perturbation incurred by the compression method can be translated into measurable noise to ensure privacy. Nevertheless, they find that quantization in itself is insufficient for privacy. The primary challenge is the exact distribution of the incurred noise (i.e., quantization error) is often intractable and \textit{dependent} on the input distribution~\cite{yan2023killing}. To solve that, \cite{yan2023killing} adds binomial noise while~\cite{youn2023randomized} to re-shape the quantization noise distribution. 

This work fills this critical gap. \textbf{Our primary contribution lies in the development of a new quantization-based distributed SGD algorithm that concurrently attains communication efficiency, protects privacy, and guarantees bounded model convergence errors}. More specifically,  each participating client in the wireless distributed learning system employs the \textbf{Gau}ssian-\textbf{L}ayered \textbf{R}andomized \textbf{Q}uantization for compression using a layered multishift coupler~\cite{wilson2000layered, mahmoud2022randomized}. By selecting the appropriate parameter, the quantization error can exactly follow the Gaussian distribution. We delve into a theoretical analysis of the quantization error statistics and characterize the privacy assurances of Gau-LQR, expressed in terms of client-level differential privacy (CLDP). Leveraging the inherent privacy attributes of the Gau-LRQ, we propose a novel algorithm, Gau-LRQ-aided local SGD (LRQ-SGD), that satisfies the prerequisite privacy by controlling the parameter of Gau-LRQ for each communication round. We theoretically illustrate the trade-offs among communication, privacy, and learning performance under this quantization paradigm.

\textbf{Our secondary contribution enhances Gau-LRQ-SGD by harnessing the privacy budget allocation}. Some works~\cite{gong2020privacy,lee2018concentrated} have observed from empirical results that we can potentially enhance model performance by dynamically adjusting the privacy budget with the training process going. Our work is the first to suggest an optimization formula to identify the optimal privacy budget allocation. Specifically, we first examine how privacy and convergence performance responds to a \textit{dynamic} parameter selection for Gau-LRQ.  We then achieve a strategic privacy budget allocation rule by minimizing convergence error at each communication round subject to the privacy budget constraint. Theoretical findings indicate that more privacy budgets should be allocated during the latter training rounds conclusion that aligns with experimental observations in~\cite{gong2020privacy,lee2018concentrated}.

\textbf{Finally, we validate our theoretical analysis through extensive experiments on staple machine learning tasks}, including MNIST, CIFAR-10, and CIFAR-100. The results demonstrate significant improvement over privacy-preserving schemes, communication-efficiency schemes, and their combination. Besides, DLRQ-SGD outperforms LRQ-SGD, demonstrating the dynamic privacy allocation can yield extra accuracy improvements for the models. On the MNIST dataset, our DLRQ-SGD reduced the communication cost by 95.3\% compared with vanilla distributed SGD but only incur an accuracy impairment of 1.19\%.

%% file: 2.Related_Work.tex
\section{Related Work}
Both communication overhead and privacy concerns have been driving forces behind the development of distributed machine learning. Several schemes have been proposed to improve the communication efficiency in gradient-based large-scale distributed learning: sparsification~\cite{shi2019distributed}, sketching~\cite{rothchild2020fetchsgd}, quantization~\cite{alistarh2017qsgd}, less frequent communication~\cite{mcmahan2017communication,zhang2020lagc}, or their combinations~\cite{basu2019qsparse,nadiradze2021asynchronous}. Privacy preservation for machine learning has been studied in distributed learning algorithms to prevent privacy leakage from the model exchange. To counter this issue, most existing efforts~\cite{abadi2016deep} applied differential privacy (DP) mechanism to add controllable noise  to raw gradients. DP provides a rigorous framework for injecting controlled noise into model parameters, thus ensuring that an attacker cannot distinguish the true parameters from the noisy ones. Not surprisingly, losses in accuracy compared to privacy-free or non-compressed gradient updates are essentially unavoidable~\cite{wei2020federated}, \cite{niknam2020federated}.





Some recent work has considered communication-efficiency and privacy-preserving abilities together in distributed learning~\cite{zong2021communication,mohammadi2021differential}, however, in a straightforward way via adding continuous noise first and then quantizing the noisy gradient. These straightforward methods usually have a high implementation complexity and lack adaptability to limited resources. Different from them, \cite{kairouz2021distributed,agarwal2021skellam,agarwal2018cpsgd} have sought to leverage various forms of discrete additive DP noise in conjunction with quantization. However, these methods do not fully exploit the privacy advantages offered by quantization itself. 

The privacy properties of compression have been observed by some early works~\cite{xiong2016randomized, jonkman2018quantisation}, which illustrate that one can translate the perturbation incurred by the compression method into measurable noise to ensure privacy. Inspired by this inherent connection, some works simultaneously achieve communication efficiency and privacy benefits by leveraging  quantization~\cite{amiri2021compressive,li2022communication}. \cite{youn2023randomized} and~\cite{yan2023killing} built a theoretical framework to theoretically prove the privacy guarantee that quantification can provide. However, their proposed quantization scheme themselves are insufficient for privacy. \cite{yan2023killing} adds additional binomial noise while~\cite{youn2023randomized} employs extra randomized sub-sampling operation, which unavoidable comprise the model utility.

Our proposed solution largely differs from existing ones in two aspects. First, our method uses one stone (i.e., quantization) to kill two birds (i.e., communication efficiency and privacy) using \textbf{only} quantization scheme with guaranteed learning performance.  Besides, we leverage a \textbf{dynamic} privacy budget allocation strategy to enhance the privacy/communication-accuracy trade-offs. The differences between our work and related literature are summarized in Table.~\ref{tab:related}.

\begin{table}[ht]
\small
	\caption{Related work comparison}
	\label{tab:related}
	\begin{tabular}{cccc}
		\hline
		Methods                                                   & Implementation                                                                  & \begin{tabular}[c]{@{}c@{}}Privacy \\ Analysis \end{tabular} & \begin{tabular}[c]{@{}c@{}}Convergence \\ Analysis\end{tabular} \\ \hline
		{\cite{zong2021communication,mohammadi2021differential}}                                                   & \begin{tabular}[c]{@{}c@{}}Quantization with\\ Continuous noise\end{tabular}    & $\times$                                                                            & $\times$                                                              \\ \hline
		\begin{tabular}[c]{@{}c@{}}QG-SGD\\ (Alg.~\ref{alg:QG-FedAvg})\end{tabular} & \begin{tabular}[c]{@{}c@{}}Quantization with\\ Continuous noise\end{tabular}    & \checkmark                                                                            & \checkmark                                                             \\ \hline
		{\cite{agarwal2018cpsgd,agarwal2021skellam,kairouz2021distributed}}                                                   & \begin{tabular}[c]{@{}c@{}}Quantization with\\ Discrete noise\end{tabular}      & \checkmark                                                                            & $\times$                                                              \\ \hline
		BQ-SGD~\cite{yan2023killing}                                                  & \begin{tabular}[c]{@{}c@{}}Quantization with\\ Discrete noise\end{tabular}      & \checkmark                                                                           & \checkmark                                                             \\ \hline
		{\cite{youn2023randomized}}                                                   & \begin{tabular}[c]{@{}c@{}}Quantization with\\ random sub-sampling\end{tabular} & \checkmark                                                                            & $\times$                                                              \\ \hline
		\begin{tabular}[c]{@{}c@{}}LRQ-SGD\\ (Alg.~\ref{alg:LRQ-SGD})\end{tabular} & Quantization                                                                    & \checkmark                                                                            & \checkmark                                                             \\ \hline
	\end{tabular}
\end{table}


%% file: 3.Problem_Formulation_new.tex
\section{Preliminaries} \label{sec:preliminaries}
\subsection{Distributed local SGD}
We consider a distributed learning problem, where $N$ clients collaboratively participate in training a shared model via a central server.
The local dataset located at client $i$ is denoted as $\mathcal{D}^{(i)}$, and the union of all local datasets $\mathcal{D} = \{\mathcal{D}^{(1)},...,\mathcal{D}^{(N)}\}$. Our goal is to find a set of global optimal parameters $\bm{\theta}$ by minimizing the objective function $F: \mathbb{R}^d \to \mathbb{R}$,
\begin{equation}
\begin{array}{llll}	\min_{\bm{\theta} \in \mathbb{R}^d}F(\bm{\theta}) = \sum_{i=1}^N  p_i \mathbb{E}_{\xi^{(i)}\sim \mathcal{D}^{(i)}}[l(\bm{\theta};\xi^{(i)})],
	\end{array}\label{optim_problem}
\end{equation}
where $p_i$ \footnote{ Note this work focuses on exploring the impact of server-client communication rather than the server-side aggregation. Our proposed method can be used in combination with different aggregation schemes, like \cite{mcmahan2017communication},\cite{blanchard2017machine},\cite{yin2018byzantine}.} is the weight of client $i$ and $l(\bm{\theta};\xi^{(i)})$ is the local loss function of the model $\theta$ towards one data sample $\xi^{(i)}$.

A standard approach to solve this problem is local SGD~\cite{zhou2018convergence} with $K$ communication rounds. Specifically, at the $k$-th round ($0\leq k \leq K-1$), the server selects a subset $\mathcal{B}_k$ of $B$ clients ($B\leq N$, each client $i$ is chosen with probability $p_i$). Each selected client $i \in \mathcal{B}_k$ downloads the global model $\bm{\theta}_k$ from server and initialize $\bm{\theta}^{(i)}_{k,0} = \bm{\theta}_k$. It then performs $Q$ local SGD to update the local model:
\begin{equation}
		\bm{\theta}^{(i)}_{k,q+1} =\bm{\theta}^{(i)}_{k,q} - \eta \bm{g}^{(i)}_{k,q}
		\label{eq:local update}
\end{equation}
for step $q = 0, 1,..., Q-1$, where $\eta$ is the local learning rate, $\bm{g}^{(i)}_{k,q}$ is local stochastic gradient based on local model parameter $\bm{\theta}^{(i)}_{k,q}$. Let $\bm{\Delta}^{(i)}_k = \bm{\theta}^{(i)}_{k,Q} - \bm{\theta}^{(i)}_{k,0}$ denotes the model update in the $k$-th communication round. Then the server aggregates model updates from  $\mathcal{B}_k$ and sends the updated global model $\bm{\theta}_{k+1}$ back to all clients for the next round's local training:
\begin{equation}
\begin{array}{llll}
    \bm{\theta}_{k+1} = \bm{\theta}_k + \cfrac{1}{B}\sum_{i \in \mathcal{B}_k}\bm{\Delta}^{(i)}_k
    \end{array}
\end{equation}
We make the following two common assumptions on such the raw gradient $\bm{g}^{(i)}_{k,q}$ and the objective function $F(\bm{\theta})$~\cite{{bottou2018optimization},{tang2019doublesqueeze}}: 
\begin{assumption}[Unbiasness and Bounded Variance of $\bm{g}^{(i)}_{k,q}$]
	For parameter $\bm{\theta}^{(i)}_{k,q}$, the stochastic gradient $\bm{g}^{(i)}_{k,q}$ is an unbiased estimator of the true gradient $\nabla F(\bm{\theta}^{(i)}_{k,q})$ with a bounded variance:
	\begin{align}
	\mathbb{E}[\bm{g}^{(i)}_{k,q}] = \nabla F(\bm{\theta}^{(i)}_{k,q}),\\
	\mathbb{E}[\|\bm{g}^{(i)}_{k,q}-\nabla F(\bm{\theta}^{(i)}_{k,q})\|^2] \le \alpha^2.
	\end{align}
	\label{ass:stochastic_gradient} 
\end{assumption}
\vspace{-5mm}
\begin{assumption}[Smoothness]
	The objective function $F(\bm{\theta})$ is $\nu$-smooth: $\forall \bm{\theta},\bm{\theta}' \in \mathbb{R}^d$, $\|\nabla F(\bm{\theta})-\nabla F(\bm{\theta}')\| \leq \nu\|\bm{\theta}-\bm{\theta}'\|$.
	\label{ass:smoothnesee} 
\end{assumption}
Assumption~\ref{ass:smoothnesee} further implies that $\forall \bm{\theta},\bm{\theta}' \in \mathbb{R}^d$, we have \begin{equation}
F(\bm{\theta}') \leq F(\bm{\theta}) + \nabla F(\bm{\theta})^\top (\bm{\theta}'-\bm{\theta}) + \frac{\nu}{2} \|\bm{\theta}'-\bm{\theta}\|^2.
\label{eq:smooth_1}
\end{equation}
In this paper, we use a weighted mean of the gradient norm over $K$ communication rounds to evaluate the learning performance, denoted as 
\begin{equation}
\mathcal{E} = \frac{\sum_{k=0}^{K-1} \tau^{-k}\|\nabla F(\bm{\theta}_k)\|^2}{\sum_{k=0}^{K-1}\tau^{-k}}    
\end{equation}
where $\tau < 1$, we give more significant weight to the gradient norm in the later stage of training, which can better capture the convergence characteristics of non-convex problems~\cite{yan2022acsgd}.

\subsection{Client-Level DP and Gaussian Mechanism}
In this work, we use client-level differential privacy (CLDP) to quantify the privacy guarantees of the proposed algorithm. Unlike data sample-level differential privacy~\cite{abadi2016deep}, which protect a single data point’s contribution in learning a model, CLDP prevents the eavesdroppers from identifying the participation of a client by observing the aggregated model update. The formal definition of $(\epsilon, \delta)$-CLDP is as follows.
\begin{definition}[$(\epsilon, \delta)$ - CLDP~\cite{geyer2017differentially, cheng2022differentially}]
	Given a set of data sets $\mathcal{D} = \{\mathcal{D}^{(1)},...,\mathcal{D}^{(N)}\}$ and a query function $q: \mathcal{D} \to \mathcal{X}$, a mechanism $\mathcal{M}: \mathcal{X} \to \mathcal{O}$ to release the answer of the query, is defined to be $(\epsilon, \delta)$ - CLDP if for any adjacent datasets $(\mathcal{D}, \mathcal{D}')$ constructed by adding or removing all records of any one client and any measurable subset outputs $O \in \mathcal{O}$,
	\begin{equation}
		\Pr\{\mathcal{M}[q(\mathcal{D})]\in O\}\leq \Pr\{\mathcal{M}[q(\mathcal{D}')]\in O\}e^{\epsilon} + \delta, \label{eq:DP}
	\end{equation}
	\label{def:CLDP}
\end{definition}
\vspace{-3mm}
\noindent where $\epsilon>0$ is the distinguishable bound of all outputs on adjacent datasets $\mathcal{D},\mathcal{D}'$ that differ in at most one client's local dataset. $\delta$ represents the event that the ratio of the probabilities for two adjacent datasets $\mathcal{D},\mathcal{D}'$ cannot be bounded by $e^\epsilon$ after privacy-preserving mechanism $\mathcal{M}$. 

The Gaussian mechanism~\cite{geyer2017differentially, cheng2022differentially} is commonly used for achieving CLDP in distributed learning by distorting the update aggregation with additive noise. It first performs \textbf{norm clipping} on each gradient update, i.e., the update $\bm{\Delta}^{(i)}_k$ is scaled by 
$\frac{\bm{\Delta}^{(i)}_k}{\max{\{1, \|\bm{\Delta}^{(i)}_k\|_2/S_2}\}}$. Scaling ensures that the $l_2$ norm is no large than $S_2$. The Gaussian mechanism then \textbf{adds noise} (scaled to $S_2$) to the sum of all scaled updates. Dividing the Gaussian mechanism's output by $B$ yields a private update aggregation:
  \begin{align}\label{eq:GM}
        \bm{\bar \Delta}_k = \frac{1}{B}\sum_{i \in \mathcal{B}_k}\Big[\cfrac{\bm{\Delta}^{(i)}_k}{\max{\{1, \|\bm{\Delta}^{(i)}_k\|_2/S_2}\}} + \mathcal{N}(0,\sigma^2\bm{I})\Big]
    \end{align}

\begin{lemma}[Moment Accountant Theorem for CLDP~\cite{cheng2022differentially}]\label{lemma:MCT}
Using Gaussian mechanism in Eq.~\eqref{eq:GM}, the private model updates $\bm{\bar \Delta}_k$ satisfied $(\frac{2S_2\sqrt{KB\ln[1/\delta]}}{N\sigma}, \delta)$-CLDP. To ensure the $\bm{\bar \Delta}_k$ satisfied $(\epsilon, \delta)$-CLDP, we need to scale the Gaussian noise level with $\sigma = \frac{2S_2\sqrt{KB\ln[1/\delta]}}{N\epsilon}$. 
\end{lemma}

\subsection{Layered Randomized Quantization}
Most methods add Gaussian noise before compressing the noisy model updates to achieve communication efficiency and privacy preservation~\cite{zong2021communication,mohammadi2021differential}. Instead, we aim to obtain privacy guarantees directly with only compression noise. Thus, the crucial problem is finding a compression technique to generate a \textit{controllable} noise distribution. Fortunately, \cite{wilson2000layered, mahmoud2022randomized} offer Layered Randomized Quantization (LRQ) based on the dithered quantization \cite{schuchman1964dither,gray1993dithered} and layered multishift coupling~\cite{wilson2000layered}, to generate exact noise distribution. 

\begin{definition}[Dithered Quantizer]\label{def:dithered}
The dithered quantizer is defined using i)  a \textit{deterministic} quantization step size $q_{step}$ and a dither signal $x \sim U(-\frac{q_{step}}{2},\frac{q_{step}}{2})$; ii) a \textit{encoder} to encode the input $u$ to $m = \Big\lfloor \frac{u + x}{q_{step}}  + \frac{1}{2}  \Big\rfloor$; and iii) a \textit{decoder} to decode $m$ to $\hat{u} =  m q_{step} - x$. 
\end{definition}

\begin{figure}[ht]
    \subfigure[Gaussian distribution]{\label{fig:GD}
		\includegraphics[width=0.29\linewidth]{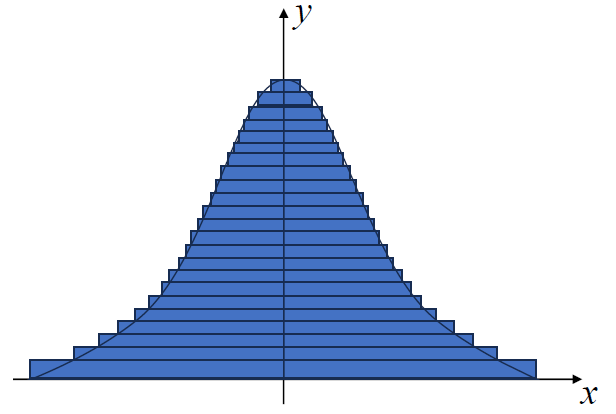}
	}
	\subfigure[Laplace distribution]{\label{fig:LD}
		\includegraphics[width=0.29\linewidth]{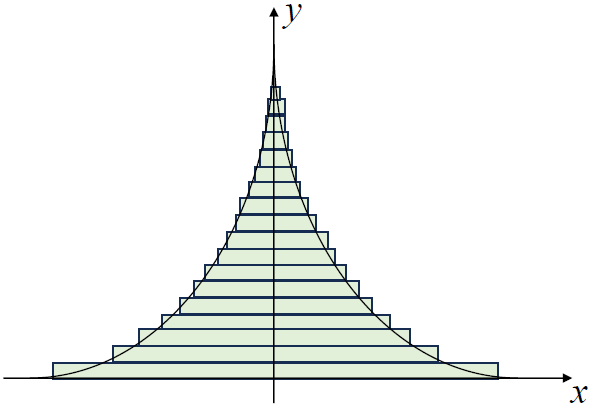}
	}
    \subfigure[Uniform distribution]{\label{fig:UD}
		\includegraphics[width=0.29\linewidth]{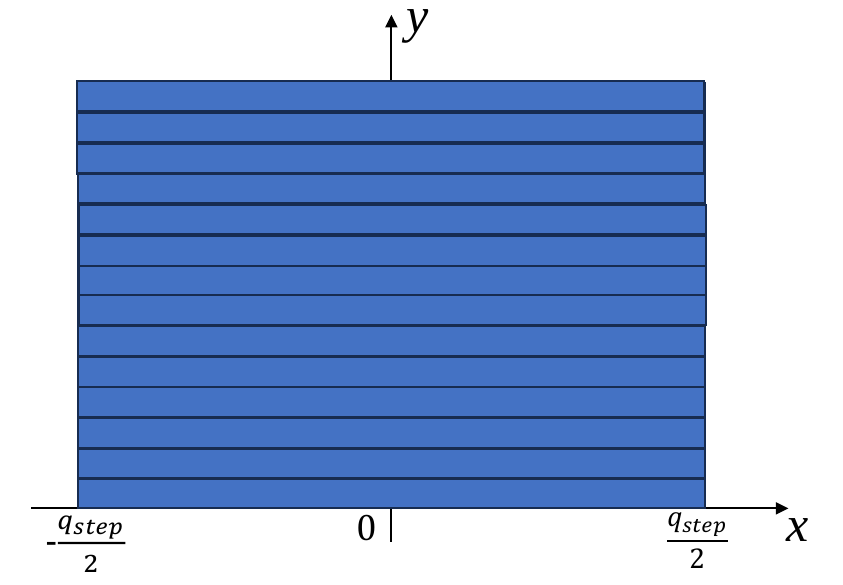}
	}
	\caption{Expressing the Gaussian and Laplace distribution as a convex combination of uniform distributions}
	\label{fig:unimodal_distribution}
\end{figure}

It can be checked that for an input $u\in [a_1,a_2]$ , the quantization noise $u - \hat{u}$ follows a uniform distribution $U(-\frac{q_{step}}{2},\frac{q_{step}}{2})$ \cite{schuchman1964dither}. Further using layered multishift coupling~\cite{wilson2000layered}, we can construct a quantizer where the quantization step $q_{step}$ is not a deterministic but random, giving a quantization noise that is a convex combination of uniform distributions. As illustrated in Fig.~\ref{fig:unimodal_distribution}, by choosing a suitable distribution of $q_{step}$, we can have any \textit{symmetric unimodal distribution} as the quantization error distribution, for example, Gaussian distribution (Fig.~\ref{fig:GD}) and Laplace distribution( Fig.~\ref{fig:LD}).

%% file: 4.LRQ-SGD_new.tex
\section{Gaussian Layered Randomized Quantized Local SGD} \label{sec:LRQ-SGD}

In this section, we first describe a new quantization mechanism, named Gaussian Layered Randomized Quantizer (Gau-LRQ) which can generate quantization noise that follows the specific Gaussian distribution. We then incorporate Gau-LRQ into a distributed local SGD framework and give the theoretical performance analysis towards privacy, communication cost, and convergence error bound.  

\subsection{Gaussian Layered Randomized Quantizer}
Instead of giving the deterministic quantization step as a dithered quantizer, we construct the Gau-LRQ with an input parameter $\sigma$, defined as: 
\begin{definition}[Gaussian Layered Randomized Quantizer] \label{definition:Gau-LRQ}
Given a parameter $\sigma$, let $x$ is randomly sampled from $\mathcal{N}(0,\sigma^2)$ and $y$ from $\exp{\{-\cfrac{(x/\sigma)^2}{2}\}}\cdot U(0,1)$. If $x <0$, then $y = 1-y$. Then the Gau-LRQ is defined using i) a \textit{random} quantization step $q_{step} = R-L$ determined by $L = -\sigma \sqrt{-2\ln{(1-y)}}$ and $R = \sigma \sqrt{-2\ln{y}}$; ii) a \textit{encoder} to encoder the input $u$ to $ m = \lfloor \cfrac{u + R + x}{R-L} \rfloor$ and iii) a \textit{decoder} to decode $m$ to $\hat{u} = m(R-L) - x$.
\end{definition}

\begin{figure}[ht]
\subfigure[Original Gaussian distribution]{\label{fig:OGD}
		\includegraphics[width=0.45\linewidth]{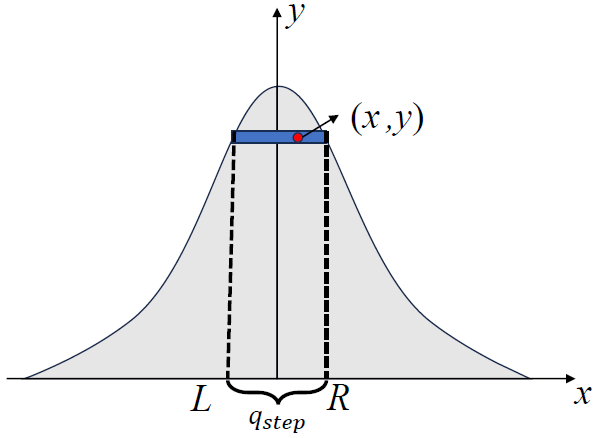}
	}
	\subfigure[Flipped Gaussian distribution]{\label{fig:FGD}
		\includegraphics[width=0.45\linewidth]{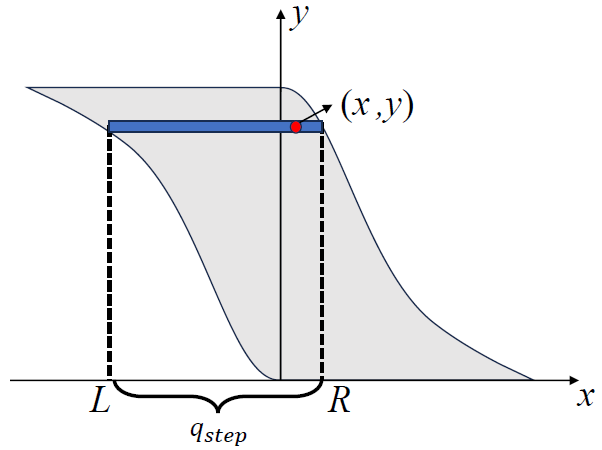}
	}
	\caption{Compute the quantization step of Gau-LRQ by sampling point $(x,y)$.}
	\label{fig:Gaussian_distribution}
\end{figure}

As shown in Fig.~\ref{fig:OGD}, we can draw $x$ (w.r.t $\sigma$) from the Gaussian distribution, determine the range of possible values for $y$, and then pick $y$ uniformly at random from within this range. To compute $L$ and $R$, we need to be able to invert the probability density function, which we can easily do for the Gaussian distribution. However, in practice, we choose to draw the $(x,y)$ from the flipped Gaussian distribution (Fig.~\ref{fig:FGD}) rather than the original Gaussian distribution (Fig.~\ref{fig:OGD}) 

There are two reasons why we consider the flipped Gaussian distribution. Firstly, in Fig.~\ref{fig:FGD}, the left region is obtained by vertically reflecting the left region of the original Gaussian distribution. Therefore, the flip operation will not affect the fact that $x$ follows the Gaussian distribution. Secondly, for Fig.~\ref{fig:OGD}, there is no lower bound on how small the $q_{step}$ can be. Quantization becomes meaningless as the value of $q_{step}$ tends to infinity. In contrast, for Fig.~\ref{fig:FGD}, there is a positive minimum length for the $q_{step}$. To characterize the privacy and communication properties of Gau-LRQ, we derive the following two lemmas.


\begin{lemma}[Quantization Noise of Gau-LRQ~\cite{wilson2000layered, mahmoud2022randomized}]
\label{lemma:LRQ_noise}
    For an input $u$, the quantization noise of Gau-LRQ $u - \hat{u}$ follows Gaussian distribution $\mathcal{N}(0,\sigma^2)$.  
\end{lemma}

\begin{lemma} [Communication Cost of Gau-LRQ]
\label{lemma:LRQ_com}
For an input signal $u \in [a_1,a_2]$, the $q_{step}$ is lower bounded by $2\sigma\sqrt{2\ln{2}}$, the quantization level $l$ is upper bounded by  $l = \frac{a_2 - a_1}{q_{step}} \le \frac{a_2 - a_1}{2\sigma\sqrt{2\ln{2}}}$, and the needed quantization bits are upper bounded by  $\log_2{[\frac{a_2 - a_1}{2\sigma\sqrt{2\ln{2}}}+1]}$.
\end{lemma}


\subsection{Gau-LRQ-SGD}
\begin{figure}
    \centering
    \includegraphics[width=\linewidth]{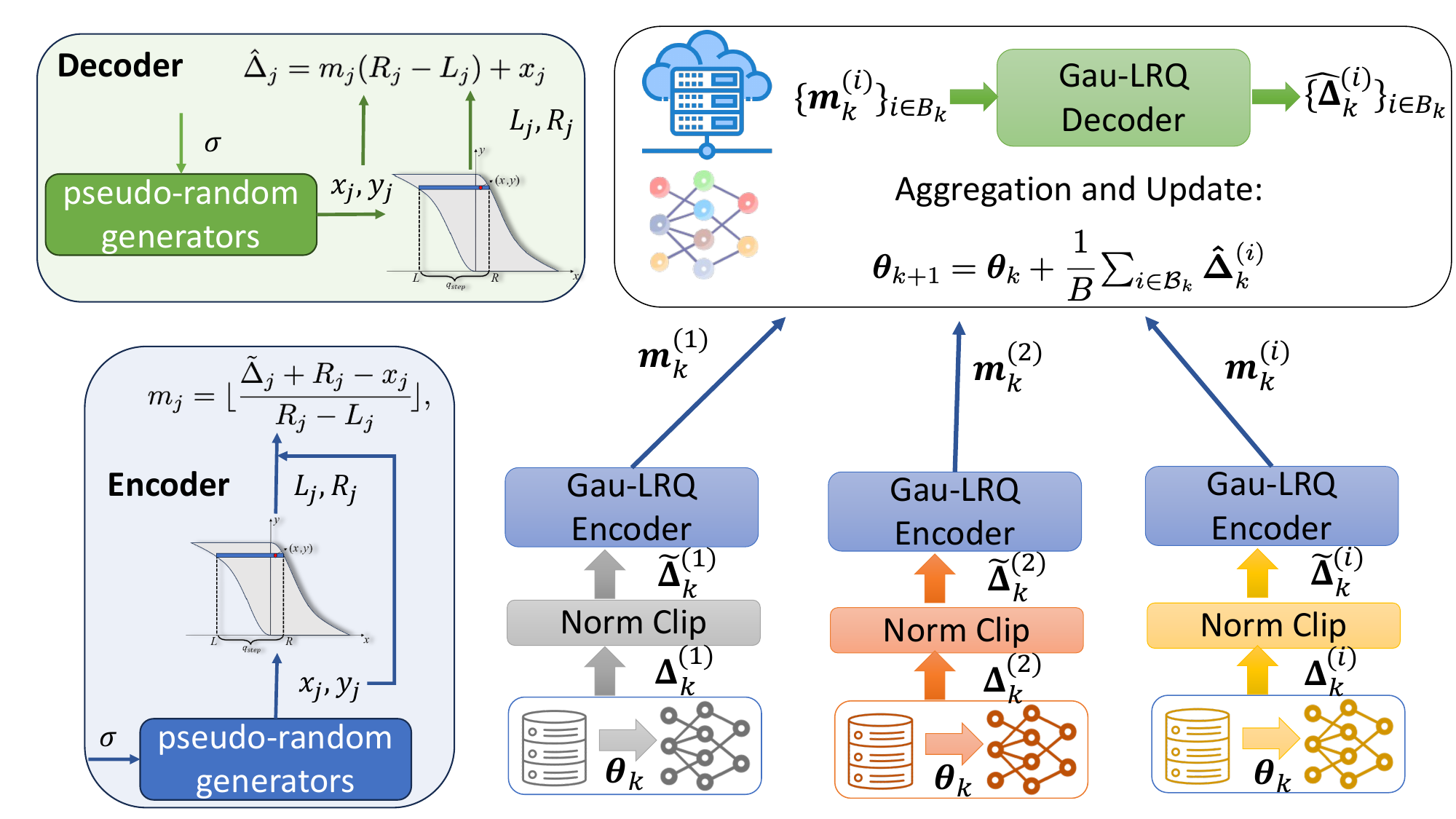}
    \caption{Gau-LRQ-SGD framework.}
    \label{fig:framework}
\end{figure}

A natural idea is to use the Gaussian quantization noise as the "perturbation" to achieve CLDP. We next incorporate the designed Gau-LRQ into the distributed SGD framework, leading to the Gau-LRQ local SGD algorithm (Alg.~\ref{alg:LRQ-SGD}), to concurrently attain communication efficiency and privacy preservation. The algorithm pipeline is summarized in Fig.~\ref{fig:framework}.

\textbf{Norm Clip}. We first clip the raw model update $\mathbf{\Delta}^{(i)}_k$ into $l_2$ norm with threshold $S_2$: 
\begin{equation}
	\Tilde{\mathbf{\Delta}}^{(i)}_k = \frac{\mathbf{\Delta}^{(i)}_k}{\max{\{1, \|\mathbf{\Delta}^{(i)}_k\|_2/S_2}\}}. \label{eq:clipupdate} 
\end{equation}
The clipped model update limits the impact of one client's samples on the whole model.

\textbf{Encode and Upload at Client $i$}. Then, each client $i$ uses the Gau-LRQ to quantize the clipped model updates in an element-wise way. In particular, the $j$-th element of $\Tilde{\mathbf{\Delta}}^{(i)}_k$, denoted as $\Tilde{\Delta}_j$ for simplicity, is encoded as:
\begin{align}\label{eq:encoder}
    m_j = \lfloor \cfrac{\Tilde{\Delta}_j+R_j-x_j}{R_j-L_j} \rfloor,
\end{align} 
where $x_j$, $L_j$ and $R_j$ are determined by Definition~\ref{definition:Gau-LRQ}. After encoding, client $i$ sends the quantized model update $\mathbf{m}_k^{(i)}$ instead of raw $\mathbf{\Delta}_k^{(i)}$ to server. 

\textbf{Decode and Aggregate at Server}. For each received quantized model update, the server decodes the $j$-th element of $\mathbf{m}_k^{(i)}$ as:
\begin{align}\label{eq:decoder}
    \hat{\Delta}_j = m_j(R_j-L_j) + x_j
\end{align}

In Definition~\ref{definition:Gau-LRQ}, it is specified that the server must know the two real numbers $x_j$ and $y_j$ to decode $\hat{\Delta}_j$ ($L_j$ and $R_j$ can be determined by $x_j$ and $y_j$.). Transmitting $x_j$ and $y_j$ directly can hinder communication efficiency. To address this issue, pseudo-random generators are utilized on both the server and clients to generate the same sequence of $x_j$ and $y_j$ synchronously. This approach only requires a one-time distribution of initial seeds to the pseudo-random generators, which can be easily achieved using existing methods such as on-demand key distribution~\cite{paladi2021ondemand} before the training process commences. Finally, the server aggregates the decoded model and updates the global model parameters.


\begin{algorithm}[ht] 
	\caption{Gaussian Mechanism-Layered Randomized Quantized Local SGD (Gau-LRQ-SGD)}
	\begin{algorithmic}[1]
	\State \textbf{Input:} Learning rate $\eta$, initial point $\bm{\theta}_0 \in \mathbb{R}^d$, local updates $Q$, communication round $K$, model update norm bound $S_2$, privacy budget $(\epsilon$, $\delta)$;
    \State Set $\sigma = \frac{2S_2\sqrt{KB\ln[1/\delta]}}{N\epsilon}$ as the parameter for Gau-LRQ;
		\For {each communication rounds $k = 0, 1, ..., K$:}
		\State \textbf{On each client {$i \in \mathcal{B}_k$}:}
		\State Download $\bm{\theta}^{(i)}_{k,0} = \bm{\theta}_{k}$ from server;
		\For {each local updates $q = 0, 1, ..., Q-1$}
		\State Performe local SGD using Eq.~(\ref{eq:local update});
		\EndFor
		\State Compute the model update $\mathbf{\Delta}^{(i)}_k = \bm{\theta}^{(i)}_{k,Q} - \bm{\theta}^{(i)}_{k,0}$;
        \State Clipping $\mathbf{\Delta}_k^{(i)}$ to $\mathbf{\Tilde{\Delta}}_k^{(i)}$ using Eq.~(\ref{eq:clipupdate});
		\State Encode $\mathbf{\Tilde{\Delta}}_k^{(i)}$ as $\mathbf{m}^{(i)}_k$ using Gau-LRQ's encoder (Eq.~\eqref{eq:encoder});
		\State Send $\mathbf{m}^{(i)}_k$ to the server;
 	\State \textbf{On the server:}
        \State Decode $\hat{\mathbf{\Delta}}_k^{(i)}$ using Gau-LRQ's decoder (Eq.~\eqref{eq:decoder});
		\State Aggregate decoded updates  $\mathbf{\bar \Delta}_k = \frac{1}{B}\sum_{i \in \mathcal{B}_k}\mathbf{\hat \Delta}^{(i)}_k$;
		\State Update global model parameter: $\bm{\theta}_{k+1} = \bm{\theta}_k + \bar{\mathbf{\Delta}}_k$;
	\EndFor
\end{algorithmic} 
	\label{alg:LRQ-SGD}
\end{algorithm}

\begin{theorem}[Performance of Gau-LRQ-SGD]
For an $N$-client distributed learning problem, Alg.~\ref{alg:LRQ-SGD} satisfies the following performance: \\	
    \textbf{Privacy:}  The LRQ-SGD satisfied $(\epsilon, \delta)$-CLDP.\\
	 \textbf{Communication:} The LRQ-SGD incurs communication cost $\sum_{k=0}^{K-1}\sum_{i=1}^B d\log_2{\Big[\frac{\|\Tilde{\mathbf{\Delta}}^{(i)}_k\|_{\infty}N\epsilon}{2S_2 \sqrt{2\ln{2}KB\ln[1/\delta]}}+1\Big]}$;\\
	 \textbf{Convergence:} The convergence error for smooth objective is upper bounded by
     \begin{align}\label{eq:cov_of_LRQ-SGD}
    &\mathcal{E} \le \underbrace{\cfrac{2[F(\bm{\theta}_0)-F(\bm{\theta}^*)]}{Q\eta\sum_{k=0}^{K-1}\tau^{-k}}+ \cfrac{Q\alpha^2}{B}\Big[\cfrac{(2Q-1)(Q-1)}{6Q}+1\Big]}_{\triangleq\mathcal{E}_{LSGD}} \nonumber\\
    &~~+\underbrace{\cfrac{4dS_2^2K\ln[1/\delta]}{\eta^2QN^2\epsilon^2}}_{\text{\rm Privacy Error}}
    \end{align}
\label{Theorem 1}
\end{theorem}

The above Theorem characterizes the privacy, communication, and convergence performance.

$\bullet$ \textbf{Privacy}. We first see there is no additional noise was added, and privacy protection is solely achieved through the quantizer. According to Lemma~\ref{lemma:LRQ_noise}, the encoder results in $\hat{\Delta}_j = \Tilde{\Delta}_j + N(0,\sigma^2)$. The quantization error, following the Gaussian distribution with $\sigma$ as standard deviation, can be viewed as a Gaussian noise added on the clipped model update. Furthermore, by controlling the value of $\sigma =  \frac{2S_2\sqrt{KB\ln[1/\delta]}}{N\epsilon}$, we can achieve the same $(\epsilon, \delta)$-CLDP of privacy protection as the Gaussian Mechanism we introduced in Section~\ref{sec:preliminaries}. The added noise variance $\sigma$ is inversely proportional to the privacy budget $\epsilon$, where a small privacy budget leads to a large $\sigma$, as well as a large quantization error (i.e., noise) on the clipped model update.

$\bullet$ \textbf{Communication}. We next assess how much communication cost is reduced by the quantization. Noted that $\Tilde{\Delta}_j \in [-\|\Tilde{\mathbf{\Delta}}^{(i)}_k\|_{\infty}, \|\Tilde{\mathbf{\Delta}}^{(i)}_k\|_{\infty}]$ and $\sigma =  \frac{2S_2\sqrt{KB\ln[1/\delta]}}{N\epsilon}$, deriving from Lemma~\ref{lemma:LRQ_com}, we can get, the quantization bits of each element is upper bounded as $b^{(i)}_k = \log_2{\Big[\frac{\|\Tilde{\mathbf{\Delta}}^{(i)}_k\|_{\infty}N\epsilon}{2S_2 \sqrt{2\ln{2}KB\ln[1/\delta]}}+1\Big]}$. $b^{(i)}_k$ is proportional to $\epsilon$, meaning that strict privacy protection can bring about lower communication costs. For simplicity, we use fixed-length coding to transmit $\mathbf{m}_k^{(i)}$, which means that each element is represented by $b^{(i)}_k$ bits. At each communication round, each client quantizes each element of the $d$-dimensional model update from $b_{init}$ to $b^{(i)}_k$ bits ($b_{init}$ is the number of bits of full-precision floating point, e.g., $b_{init}=32$ or $b_{init}=64$). We can achieve this communication cost by summing over $B$ clients, and $K$ communication rounds. Compared with vanilla local SGD, we can reduce $BKdb_{init}-\sum_{k=0}^{K-1}\sum_{i=1}^B db^{(i)}_k$ bits communication overhead.


$\bullet$ \textbf{Convergence error bound}. We finally show how the Gau-LRQ affects the learning convergence. The first item in Eq.~\ref{eq:cov_of_LRQ-SGD}, denoted as $\mathcal{E}_{LSGD}$, refers to the convergence error bound of vanill local SGD with \textit{non-compressed} and \textit{privacy-free} model updates. The second item is the privacy error, indicating the trade-off between privacy and the accuracy of our algorithm. The privacy error is inversely proportional to $1/\epsilon^2$, meaning that strict privacy protection can lessen the model's accuracy. And more communication rounds $K$ lead to more significant privacy error. This is because more communication rounds cause less privacy budget at each
round, and larger noise must be added to the model update, thus degrading the model's accuracy. The complete proof can be found in Appendix~\ref{proof:The1}. 

%% file: 5.DLRQ-SGD.tex
\section{Dynamic Layered Randomized Quantized Local SGD} \label{sec:DLRQ-SGD}
In Gau-LRQ-SGD, we divide the privacy budget $\epsilon$ equally among $K$ communication rounds. Therefore, the $\sigma$ for each round remains fixed during training. Recently, \cite{gong2020privacy,lee2018concentrated} revealed from the experimental observation that in the early stages of training, random models are far from optimal, and the gradients are usually large, which makes it possible to allocate less privacy budget (i.e., add more noise). Therefore, we can potentially enhance model convergence performance by dynamically adjusting the privacy budget in response to the current learning process. In this section, we aim to find a dynamic privacy budget allocation scheme to enhance convergence performance from a theoretic perspective. 

Instead of taking the $\sigma$ as a fixed parameter for Gau-LRQ, we use a set of noise variance $\{\sigma_0,...,\sigma_{K-1}\}$ for communication round $k$, leading to a dynamic version of Gau-LRQ. The following two lemmas show how the dynamic noise variance affects the privacy and convergence performance 

\begin{lemma}[Dynamic Moment Accountant]
\label{lemma:Dynamic_MAT}
With a set of $\{\sigma_0,...,\sigma_{K-1}\}$, the Gau-LRQ-SGD satisfied $(\epsilon', \delta)$-CLDP with:
    \begin{align}
        \epsilon' = \frac{2S_2\sqrt{B\ln[1/\delta]}}{N}\cdot\sqrt{\sum_{k=0}^{K-1}\frac{1}{\sigma_k^2}}
    \end{align}
\end{lemma}


\begin{lemma}\label{lemma:Dynamic_convergence}
With a set of $\{\sigma_0,...,\sigma_{K-1}\}$, the convergence error of Gau-LRQ-SGD for the smooth objective is upper bounded by
\begin{align}\label{eq:dynamic_privacy_error}
\mathcal{E} \le \mathcal{E}_{LSGD} +\underbrace{\frac{d}{B\eta^2Q\sum_{i=0}^{K-1}\tau^{-i}}\sum_{k=0}^{K-1}\tau^{-k}\sigma_k^2}_{\text{\rm Privacy Error}}
\end{align}
\end{lemma}


The full proof of Lemma~\ref{lemma:Dynamic_MAT} and Lemma~\ref{lemma:Dynamic_convergence} are shown in Appendix~\ref{proof:lemma5} and \ref{proof:The1}. If we fix $\sigma_k = \sigma$ for each round $k$, Lemma~\ref{lemma:Dynamic_MAT} degrades to Lemma~\ref{lemma:MCT} with $\epsilon' = \frac{2S_2\sqrt{KB\ln[1/\delta]}}{N\sigma}$; and Eq.~\eqref{eq:dynamic_privacy_error} degrades to Eq.~\eqref{eq:cov_of_LRQ-SGD}. 

To improve the convergence performance of Alg.~\ref{alg:LRQ-SGD} with a dynamic Gau-LRQ, we need to determine how to allocate the privacy budget for each round, i.e., how to determine $\sigma_k$. To solve that, we formulate the privacy budget allocation problem as a \textit{convergence error minimization} problem under total privacy budget constraints $(\epsilon,\delta)$. By inspecting Eq.~\eqref{eq:dynamic_privacy_error}, we can see that to minimize the convergence error, we need to minimize the privacy error. Therefore, we formulate 
\begin{equation}
	\begin{split}
		& \min_{\{\sigma_k^2\}} ~ \sum_{k=0}^{K-1}\tau^{-k}\sigma_k^2\\
		& s.t. ~~ \frac{2S_2\sqrt{B\ln[1/\delta]}}{N}\cdot\sqrt{\sum_{k=0}^{K-1}\frac{1}{\sigma_k^2}} = \epsilon,
		\label{eq:optimization-non}
	\end{split}
\end{equation}
By solving the above optimization problem, we can determine the vector $\{\sigma_k^2\}_{k=0}^{K-1}$ for every communication round:
\begin{equation}
	{\boxed {\sigma_k^2 = \frac{4S_2^2B\ln[1/\delta]}{N^2\epsilon^2} (\sum_{i=0}^{K-1}\tau^{-i/2})\tau^{k/2} }}
	\label{eq:dynamic_noise}
\end{equation}
Then, we can modify Alg.~\ref{alg:LRQ-SGD} to a dynamic version by determining $\sigma_k$ for the Gau-LRQ at the beginning of each communication round. Note that, if we take $\tau=1$, then $\sigma_k^2$ will be fixed as $\frac{4S_2^2KB\ln[1/\delta]}{N^2\epsilon^2}$ for all $K$ rounds, degrading to a fixed Gau-LRQ quantizer. If we consider a more general case with $\tau<1$, we have 
\begin{equation} \label{eq:dynamic_sigma}
	\sigma_k^2 = \frac{4S_2^2B\ln[1/\delta]}{N^2\epsilon^2} \cfrac{\tau^{-K/2}-1}{\tau^{-1/2}-1}\tau^{k/2}
\end{equation}

Three factors determine the $\sigma_k$: (i) the total privacy budget $(\epsilon, \delta)$, less privacy budget requires setting larger $\sigma_k$ at each round; (ii) the total number of communication rounds $K$, more communication rounds causes less privacy budget at each round, and larger $\sigma_k$; (iii) the current communication round $k$, $\sigma_k$ decreases as the training process goes on. This approach aligns with the conclusion reached by~\cite{yan2022acsgd}, which suggests that smaller compression ratios should be used during early training stages and gradually increased to minimize the convergence error.


\begin{theorem}[Performance of Dynamic Gau-LRQ-SGD]
For an $N$-client distributed learning problem in Eq.~\eqref{optim_problem}, given the total privacy budget $(\epsilon, \delta)$, clipping norm bound $S_2$, local updates $Q$, and communication rounds $K$, then Dynamic LRQ-SGD satisfies the following. \\		
\textbf{Privacy:}  Dynamic Gau-LRQ-SGD is $(\epsilon, \delta)$-CLDP;\\
 \textbf{Communication:} Dynamic Gau-LRQ-SGD incurs communication cost $d\sum_{k=0}^{K-1}\sum_{i=1}^B\log_2{\Big[\frac{\|\Tilde{\mathbf{\Delta}}^{(i)}_k\|_{\infty}N\epsilon\tau^{-k/4}}{2S_2 \sqrt{2\ln{2}KB\ln[1/\delta]\sum_{i=0}^{K-1}\tau^{-i/2}}}+1\Big]}$;\\
 \textbf{Convergence:} The convergence error for the smooth objective is upper bounded by
\begin{align}\label{eq:dyna_convergence}
   \mathcal{E}\le \mathcal{E}_{LSGD} + \cfrac{4dS_2^2K\ln[1/\delta]}{\eta^2QN^2\epsilon^2}\cdot\cfrac{AM_K^2(\tau^{-k/2})}{QM_K^2(\tau^{-k/2})}
\end{align}
where $AM_K(\tau^{-k/2}) =\frac{1}{K} \sum_{k=0}^{K-1} \tau^{-k/2}$ is the Arithmetic Mean, and $QM_K(\tau^{-k/2}) =\sqrt{\frac{1}{K} \sum_{k=0}^{K-1} \tau^{-k}}$ is the Quadratic Mean.
\label{Theorem 2}
\end{theorem}

We derive several observations from the above Theorem.

$\bullet$ \textbf{Privacy}. Take $\sigma_k$ in Eq.~\ref{eq:dynamic_noise} into Lemma~\ref{lemma:Dynamic_MAT}, we find that the private model update at round $k$ satisfied $(\epsilon_k,\delta)$-CLDP with
    \begin{align}
        \epsilon_k = \frac{2S_2\sqrt{B\ln[1/\delta]}}{N}\cdot\sqrt{\frac{1}{\sigma_k^2}} = \epsilon \sqrt{\cfrac{1}{\sum_{i=0}^{K-1}\tau^{-i/2}}}\cdot\tau^{-k/4}
    \end{align}
We can see that $\epsilon_k$ increases as the training process goes on, which means we allocate less privacy
budget (i.e., larger $\sigma_k^2$) at the early
stage of training and increase the privacy budget as the training goes on. This is consistent with the conclusion drawn by~\cite{gong2020privacy,lee2018concentrated} through a heuristic algorithm.

$\bullet$ \textbf{Communication}. When using Gau-LRQ, both the quantization step and the corresponding quantization bits change across communication rounds. Similar to Theorem~\ref{Theorem 1},the quantization bits of each element is upper bounded as $b^{(i)}_k = \log_2{\Big[\frac{\|\Tilde{\mathbf{\Delta}}^{(i)}_k\|_{\infty}N\epsilon\tau^{-k/4}}{2S_2 \sqrt{2\ln{2}KB\ln[1/\delta]\sum_{i=0}^{K-1}\tau^{-i/2}}}+1\Big]}$. We still use fixed-length coding to transmit $\mathbf{m}_k^{(i)}$, which means that each element is represented by $b^{(i)}_k$ bits. To calculate the total communication cost of Dynamic Gau-LRQ-SGD, we sum over $d$ dimensional elements in each quantized model update, $B$ clients, and $K$ communication rounds.

$\bullet$ \textbf{Convergence}. The convergence error bound has an analogy format with that in Theorem~\ref{Theorem 1}. The first item is generated by full-communication and privacy-free local SGD. In comparing the privacy error between Eq.~\eqref{eq:dyna_convergence} and Eq.~\eqref{eq:cov_of_LRQ-SGD}, we can observe that the latter is multiplied by an additional factor of $\cfrac{AM_K^2(\tau^{-k/2})}{QM_K^2(\tau^{-k/2})}$. It is worth noting that $0 <\tau <1$, which means that $AM_K(\tau^{-k/2})$ is always smaller than $QM_K(\tau^{-k/2})$. This suggests that dynamic Gau-LRQ-SGD can achieve lower convergence error when compared to fixed LRQ-SGD.The proof is shown in Appendix~\ref{proof:The1}.

%% file: 6.Discussion.tex
\section{Discussion}

From Table~\ref{tab:related}, we observe that the related literature can be classified into two lines, namely, quantization with continuous noise~\cite{zong2021communication,mohammadi2021differential} and with discrete noise~\cite{agarwal2018cpsgd,agarwal2021skellam,yan2023killing}. The subsequent discussion delineates the limitations of these approaches with respect to their convergence and privacy performance.

\subsection{Quantization with Continuous noise}
\cite{zong2021communication,mohammadi2021differential} do not give explicit theoretical performance analysis. For comparison, we develop and analyze a naive algorithm following the same quantization-post-continuous noise procedure as \cite{zong2021communication,mohammadi2021differential}, termed as Quantized Gaussian mechanism SGD (QG-SGD). Like Gau-LRQ-SGD, QG-SGD firstly clips the raw model update $\mathbf{\Delta}^{(i)}_k$ using Eq.~(\ref{eq:clipupdate}). To fulfill the privacy budget, we add Gaussian noise sampled from $\mathcal{N}(0,\sigma^2=\frac{4S_2^2KB\ln(1/\delta)}{N^2\epsilon^2})$ on the clipped model update to achieve $(\epsilon,\delta)$-CLDP. After that, we quantize each private model update to $b=\log_2{\Big[\frac{\|\Tilde{\mathbf{\Delta}}^{(i)}_k\|_{\infty}N\epsilon}{2S_2 \sqrt{2\ln{2}KB\ln[1/\delta]}}+1\Big]}$ bits by using a stochastic quantizer $\mathcal{Q}_b[\cdot]$~\cite{alistarh2017qsgd} to meet the communication constraint. For the detailed algorithm, see Alg.~\ref{alg:QG-FedAvg}. 

\begin{algorithm}[htbp] 
	\caption{Quantized Gaussian mechanism SGD (QG-SGD)}
	\begin{algorithmic}[1]
		\State \textbf{Input:} Learning rate $\eta$, initial point $\bm{\theta}_0 \in \mathbb{R}^d$, model update norm bound $S_2$ $S_{\infty}$; clipping norm bound $S_2$ and $S_{\infty}$, number of local updates $Q$, communication rounds $K$; privacy budget $(\epsilon$, $\delta)$ and communication requirement $b$;
		\For {each communication rounds $k = 0, 1, ..., K$:}
		\State \textbf{On each client {$i \in \mathcal{B}_k$}:}
		\State Download $\bm{\theta}^{(i)}_{k,0} = \bm{\theta}_{k}$ from server;
		\For {each local updates $q = 0, 1, ..., Q-1$}
		\State Performe local SGD using Eq.~(\ref{eq:local update});
		\EndFor
		\State Compute the model update $\mathbf{\Delta}^{(i)}_k = \bm{\theta}^{(i)}_{k,Q} - \bm{\theta}^{(i)}_{k,0}$;
		\State Clipping $\mathbf{\Delta}_k^{(i)}$ to $\mathbf{\Tilde{\Delta}}_k^{(i)}$ using Eq.~(\ref{eq:clipupdate});
		\State Add noise $\bm{n}^{(i)}_k$ sampled form $\mathcal{N}(0,\frac{4S_2^2KB\ln(1/\delta)}{N^2\epsilon^2})$ to $\Tilde{\mathbf{\Delta}}^{(i)}_k $;
		\State Quantize private model update $\hat{\mathbf{\Delta}}^{(i)}_k = \mathcal{Q}_b[\Tilde{\mathbf{\Delta}}^{(i)}_k +  \bm{n}^{(i)}_k]$ using stochastic quantizer~\cite{alistarh2017qsgd};
		\State Send $\hat{\mathbf{\Delta}}^{(i)}_k$ to the server;
		\State \textbf{On the server:}
		\State Decode $\hat{\mathbf{\Delta}}_k^{(i)}$ using stochastic quantizer~\cite{alistarh2017qsgd};
		\State Aggregate model updates: $\mathbf{\bar \Delta}_k = \cfrac{1}{B}\sum_{i \in \mathcal{B}_k}\mathbf{\hat \Delta}^{(i)}_k$;
		\State Update global model : $\bm{\theta}_{k+1} = \bm{\theta}_k + \bar{\mathbf{\Delta}}_k$;
		\EndFor
	\end{algorithmic} 
	\label{alg:QG-FedAvg}
\end{algorithm}

\begin{lemma}[Performance of QG-SGD]
For an $N$-client distributed learning problem, Alg.~\ref{alg:QG-FedAvg} satisfies the following performance: \\	
    \textbf{Privacy:}  The QG-SGD satisfied $(\epsilon, \delta)$-CLDP.\\
	 \textbf{Communication:} The QG-SGD incurs communication cost $\sum_{k=0}^{K-1}\sum_{i=1}^B d\log_2{\Big[\frac{\|\Tilde{\mathbf{\Delta}}^{(i)}_k\|_{\infty}N\epsilon}{2S_2 \sqrt{2\ln{2}KB\ln[1/\delta]}}+1\Big]}$;\\
	 \textbf{Convergence:} The convergence error for smooth objective is upper bounded by
     \begin{align} \label{eq:cov_of_QG-SGD} 
  \mathcal{E} &\le \mathcal{E}_{LSGD} + \underbrace{\frac{4dS_2^2K\ln[1/\delta]}{\eta^2QN^2\epsilon^2}}_{\text{\rm Privacy Error}} + \underbrace{\frac{8\ln{2}dS_2^2K\ln(1/\delta)}{\eta^2QN^2\epsilon^2}}_{\text{\rm Quantization Error}}\nonumber\\    &~~~~~+\underbrace{\frac{32\ln{2}dS_2^4K^2B\ln^2(1/\delta)}{N^4\epsilon^4\|\Tilde{\mathbf{\Delta}}\|_{\infty}^2\eta^2Q}}_{\text{\rm Coupling Error}}, 
	\end{align}
\label{lemma:QG-SGD}
\end{lemma}

Given the same privacy and communication constraints as Gau-LRQ-SGD, we observe the convergence characteristics of QG-SGD. We find the $\mathcal{E}_{LSGD}$ and the privacy error in Eq.~\eqref{eq:cov_of_QG-SGD} are the same as those in Eq.~\eqref{eq:cov_of_LRQ-SGD}. There exist extra quantization error, determined by the stochastic quantizer $\mathcal{Q}_b[\cdot]$, and the coupling error between quantization and privacy. Since our LRQ-SGD realizes communication reduction and privacy protection simultaneously \textbf{only} through quantization operation, the quantization error does not appear in Eq.~\eqref{eq:cov_of_LRQ-SGD}, which means a smaller convergence error. Although the idea of combining quantization and continuous noise is not novel, our work is the first to analyze the convergence of such algorithms and to quantify the impact of noise and quantization bits. The full proof is shown in Appendix~\ref{proof:lemma6}.

\subsection{Quantization with discrete noise}
Like our starting point, \cite{yan2023killing} attempts to use quantization to solve both privacy and communication overhead. The authors first exploit the privacy advantages offered by uniform quantization itself. They found that the quantization itself is insufficient for privacy and thereby adding additive binomial noise to reach the desired DP guarantee. The convergence performance is as follows:

\begin{lemma}[Performance of BQ-SGD~\cite{yan2023killing}]
For an $N$-client distributed learning problem, BQ-SGD satisfies the following performance: \\	
    \textbf{Privacy:}  The BQ-SGD satisfied $(\epsilon, \delta)$-CLDP.\\
	 \textbf{Communication:} The BQ-SGD incurs communication cost $\sum_{k=0}^{K-1}\sum_{i=1}^B d\log_2{\Big[\frac{\|\Tilde{\mathbf{\Delta}}^{(i)}_k\|_{\infty}N\epsilon}{2S_2 \sqrt{2\ln{2}KB\ln[1/\delta]}}+1\Big]}$;\\
	 \textbf{Convergence:} The convergence error of BQ-SGD for smooth objective is upper bounded by
     \begin{align} \label{eq:cov_of_BQ-SGD} 
        \mathcal{E}\le \mathcal{E}_{LSGD} + \underbrace{\cfrac{20d^2S_2^2K}{\eta^2QN^2\epsilon^2\delta^2}}_{\text{\rm Privacy Error}} +\underbrace{\frac{2\ln{2}dS_2^2K\ln(1/\delta)}{\eta^2QN^2\epsilon^2}}_{\text{\rm Quantization Error}}, 
	\end{align}
\end{lemma} \label{lemma7}

Similar to Lemma~\ref{lemma:QG-SGD}, given the same privacy and communication constraints as Gau-LRQ-SGD, we observe the convergence characteristics of BQ-SGD. Compared with Eq.~\eqref{eq:cov_of_BQ-SGD} and Eq.~\eqref{eq:cov_of_LRQ-SGD}, the privacy error of our Gau-LRQ-SGD is $O(\frac{d}{\epsilon^2})$ and that of~\cite{yan2023killing} is $O(\frac{d^2}{\epsilon^2})$, which means Gau-LRQ-SGD can achieve lower convergence error. The full proof is shown in Appendix.~\ref{proof:lemma7}

To sum up, in order to meet the same privacy budget, quantization with continuous noise completely separates the operations of compression and noise injection, leading to both quantization and privacy error. Although the quantization with the discrete noise method explores the limited privacy guarantees of the quantizer itself, the additional discrete noise causes a larger privacy error than our LRQ. For more experimental performance comparisons, please refer to section~\ref{sec:experiments}.



%% file: 7.Experiments.tex
\section{Experiments}
\label{sec:experiments}

In this section, we conduct experiments on MNIST, CIFAR-10 and CIFAR-100 to empirically validate our proposed Gau-LRQ-SGD and Dynamic Gau-LRQ-SGD methods. The MNIST consists of 70000 $1 \times 28 \times 28$ grayscale images in 10 classes. The CIFAR-10 dataset consists of 60000 $3 \times 32 \times 32$ color images in 10 classes, and the CIFAR-100 dataset consists of 60000 $3 \times 32 \times 32$ color images in 100 classes. We compare our proposed methods with the following baselines: 1) \textbf{Gau-SGD}~\cite{geyer2017differentially}: adds Gaussian DP noise sampled from $\mathcal{N}(0,\sigma^2=\frac{4S_2^2KB\ln(1/\delta)}{N^2\epsilon^2})$ to achieve $(\epsilon,\delta)$-CLDP and then sends the private model update using the full-precision floating-point $b_{init}=32$ to the server. Gau-SGD only considers privacy protection; 2) \textbf{QG-SGD}: adds Gaussian noise sampled from $\mathcal{N}(0,\sigma^2=\frac{4S_2^2KB\ln(1/\delta)}{N^2\epsilon^2})$ to achieve $(\epsilon,\delta)$-CLDP and then quantizes the private model update to $b^*$ bits;  3) \textbf{BQ-SGD}~\cite{yan2023killing}: first quantize the clipped updates into $s$ level using stochastic quantization~\cite{alistarh2017qsgd}, and then add an extra noise $\mathbf{o}$, which are i.i.d sampled from a Binomial distribution $Bin(m,\frac{1}{2})$, on the values after stochastic quantization, where $s$ and $m$ are determined by privacy and communication constrain; 4) \textbf{Local SGD}: as the oracle, clients send noise-free and non-compressed model updates to server. For brevity, in the legends of all the following figures, we abbreviate Gau-LRQ-SGD as LRQ-SGD and Dynamic Gau-LRQ-SGD as DLRQ-SGD.

\textbf{Experimental Setting.} We select the momentum SGD as an optimizer, where the momentum is set to 0.9, and weight decay is set to 0.0005. On CIFAR-10 and CIFAR-100, all algorithms run in Local SGD fashion at the first 10 epochs as a warm-up. Following the setup of~\cite{zhang2022understand, geyer2017differentially}, the samples on one client can overlap with those on the other clients, and the samples on each client are uniformly distributed in 10 classes. Following a procedure proposed by~\cite{geyer2017differentially}, in each communication round, we calculate the median norm of all unclipped updates and use this as the clipping bound $S_2 = \text{median}\{\|\Delta^{(i)}_k\|_2, i \in \mathcal{B}_k\}$. Following the setup of~\cite{yan2022acsgd}, we estimate $\tau$ as $\tau_{est} = \left[\frac{F(\bm{\theta}_k)}{F(\bm{\theta}_0)}\right]^{1/k}$.  Other experimental details are given in Table~\ref{tab:parameter}.
\begin{table}[ht]
	\caption{Experiment Setting.}
 \centering
	\label{tab:parameter}
\setlength{\tabcolsep}{1mm}{
				\begin{tabular}{lccccr}
					\hline
		Dataset & MNIST & CIFAR-10 & CIFAR-100\\ 
		\hline
		Net & LeNet & Resnet 18 & Resnet 34\\ 
        Model Size & $6\times 10^4$  & $1\times 10^7$& $3\times 10^7$\\
		Learning Rate & 0.01  & 0.01 & 0.01\\ 
		Batch Size    & 32 & 32  & 32\\ 
		Number of Clients       & 1920  & 9600 & 48000 \\
        Local Data size         &  500  & 800  & 800\\
        Participated Clients       & 80  & 80  & 80\\
        Decay Factor $\tau$       & 0.9  & 0.98  & 0.985\\
\hline
\end{tabular}}
\end{table}

\begin{figure*}[!htbp]
\centering
	\subfigure[MNIST]{
		\includegraphics[width=0.55\columnwidth]{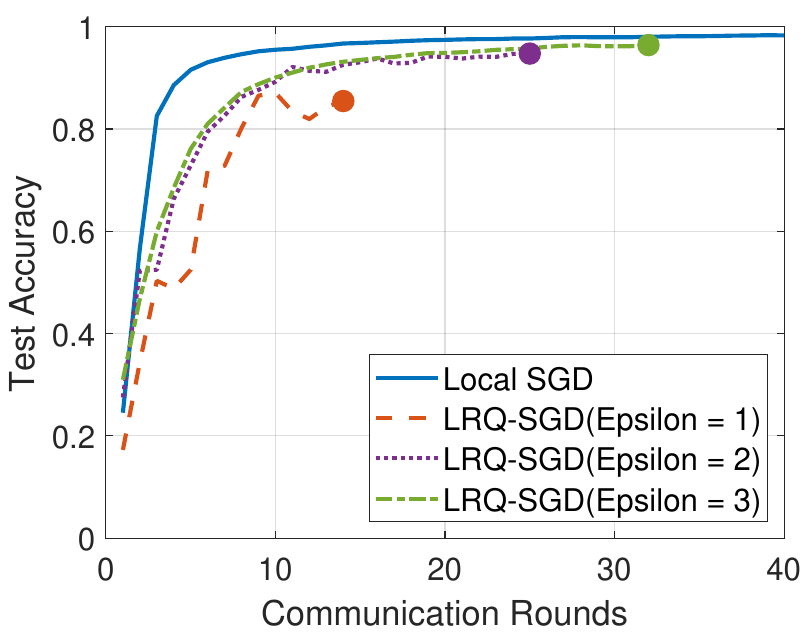}
	}
	\subfigure[CIFAR-10]{
		\includegraphics[width=0.55\columnwidth]{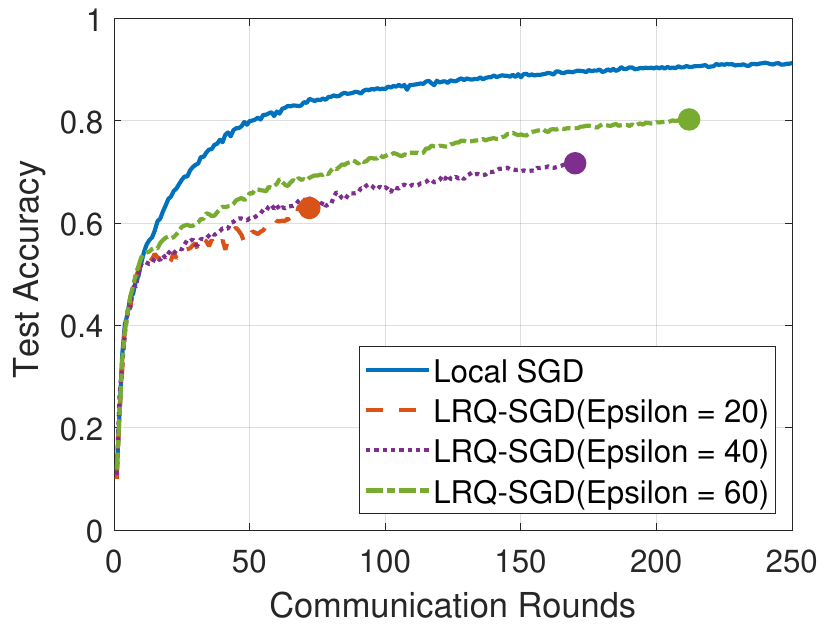}
	}
	\subfigure[CIFAR-100]{
		\includegraphics[width=0.55\columnwidth]{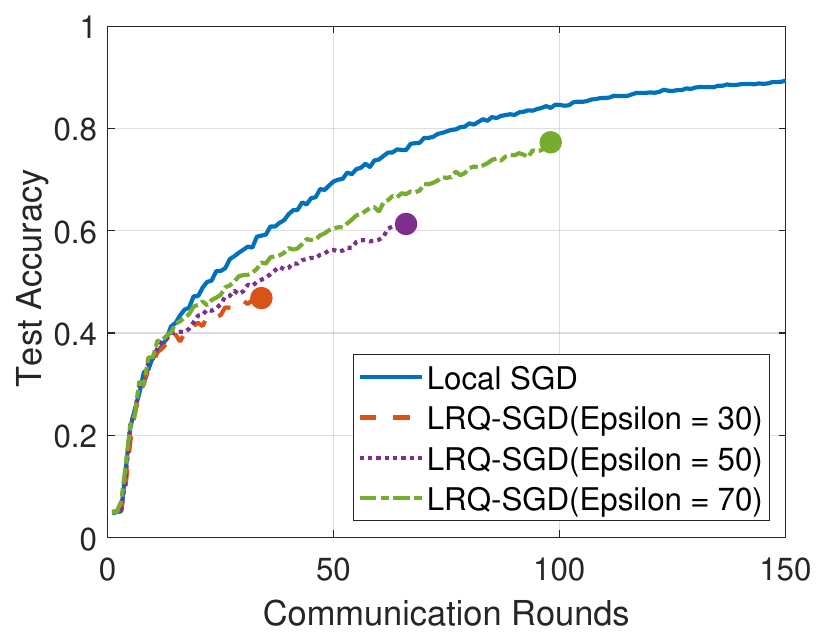}
	}
	\caption{Model Performance on Different Datasets. (Dots at the end of accuracy curves indicate that the $\delta$ threshold was reached and training therefore stopped.)}
	\label{fig:Testing Performance}
\end{figure*}

\begin{figure*}[htbp]
\centering
	\subfigure[MNIST]{
		\includegraphics[width=0.55\columnwidth]{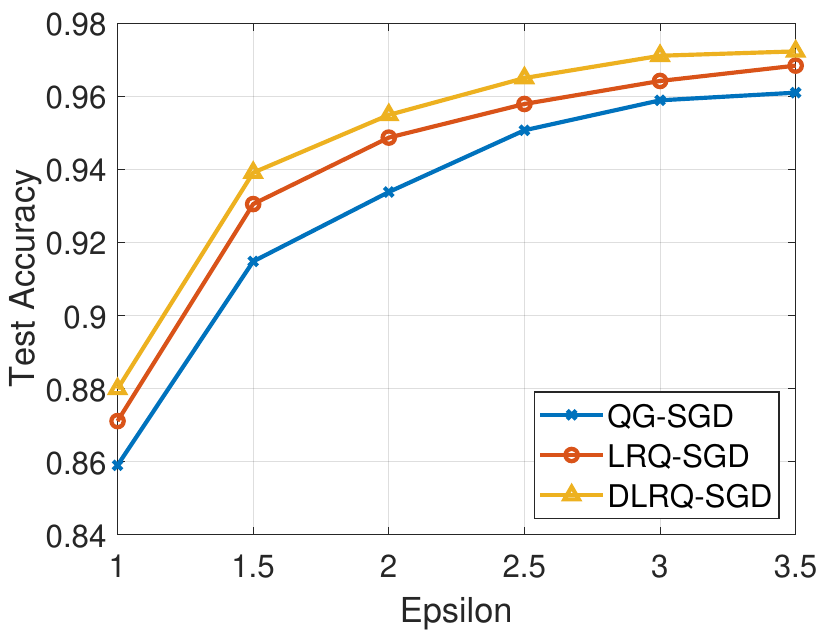}
	}
	\subfigure[CIFAR-10]{
		\includegraphics[width=0.55\columnwidth]{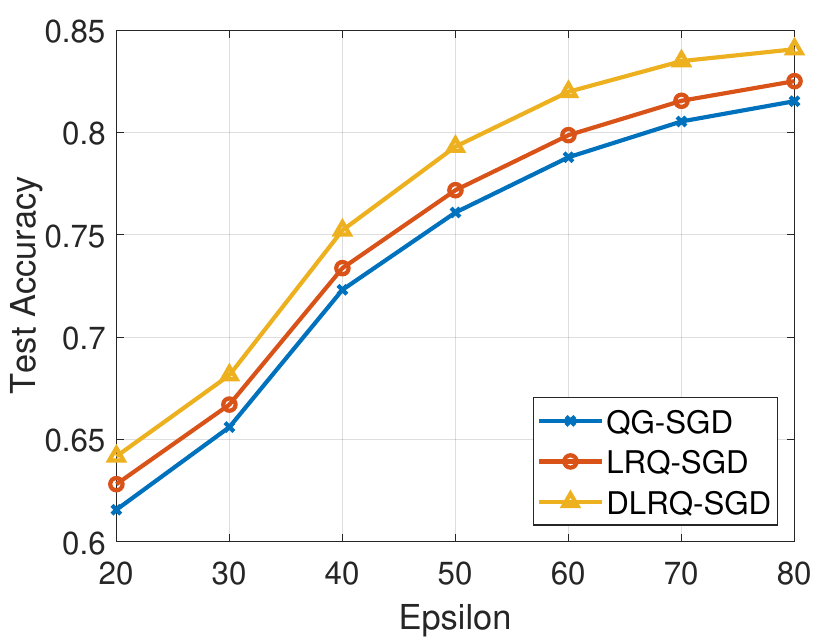}
	}
	\subfigure[CIFAR-100]{
		\includegraphics[width=0.55\columnwidth]{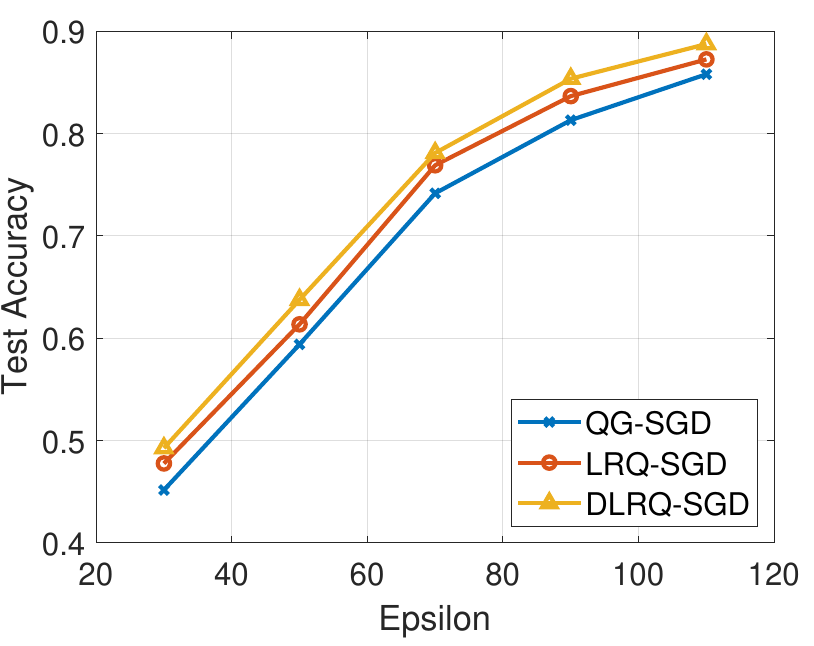}
	}
	\caption{Privacy-Learning Tradeoff on Different Datasets.}
	\label{fig:Privacy-Learning Tradeoff}
\end{figure*}

\textbf{Testing Performance.} Figure~\ref{fig:Testing Performance} shows the test accuracy of Gau-LRQ-SGD for different privacy constraints on MNIST, CIFAR-10, and CIFAR-100. For MNIST, the Local SGD can achieve a test accuracy of 0.9830 without privacy protection and incur a communication cost of 768 MB. Then we set $\delta = 1 \times 10^{-5}$ and $\epsilon = 1, 2, 3$, and we keep track of privacy loss using the privacy accountant, and training is stopped once $\delta$ reaches $1 \times 10^{-5}$. The Gau-LRQ-SGD can achieve the test accuracy of 0.8711, 0.9487, and 0.9642, respectively, and incurs the communication cost of 16.8 MB, 30 MB, and 36 MB, respectively. We can see that with the relaxation of the privacy budget (i.e., an increase in $\epsilon$), the added noise decreases, the number of communication rounds increases, and the accuracy of model testing accuracy improves. However, the corresponding communication cost will increase. Similar results can be seen in CIFAR-10 and CIFAR-100. For CIFAR-10, the Local SGD can achieve a test accuracy of 0.9215 without privacy protection and incur the communication cost of 1280 GB. Then we set $\delta = 1 \times 10^{-5}$ and $\epsilon = 20, 40, 60$, and the Gau-LRQ-SGD can achieve the test accuracy of 0.6281, 0.7337, and 0.7987, respectively. And Gau-LRQ-SGD incurs the communication cost of 43.2 GB, 60 GB, and 127.2 GB, respectively. For CIFAR-100, the Local SGD can achieve a test accuracy of 0.9207 without privacy protection and incur the communication cost of 2400 GB. Then we set $\delta = 1 \times 10^{-5}$ and $\epsilon = 30, 50, 70$, and the Gau-LRQ-SGD can achieve the test accuracy of 0.4776, 0.6136 and 0.7691, respectively. And Gau-LRQ-SGD incurs the communication cost of 81.6 GB, 158.4 GB, and 235.2 GB, respectively.

\textbf{Privacy-Learning Tradeoff.} Figure~\ref{fig:Privacy-Learning Tradeoff} shows the tradeoff between the privacy budget and the learning performance in terms of test accuracy on different datasets. We compare this tradeoff between our proposed two algorithms and the naive algorithm, QG-SGD. All three algorithms show a privacy-learning tradeoff; that is, the more privacy budget can be used, the higher test accuracy can be achieved. The marginal utility (how much test accuracy is improved from the increased communication budget) is diminishing. That means when the privacy budget is small, increasing the privacy budget can bring significant improvement. When the privacy budget is large (for example, $\epsilon>1.5$ in MNIST), the improvement of the test accuracy by increasing the privacy budget is limited.

\begin{figure}[htbp]
\centering
	\subfigure[MNIST, $(3, 1 \times 10^{-5})$-CLDP.]{
		\includegraphics[width=0.46\columnwidth]{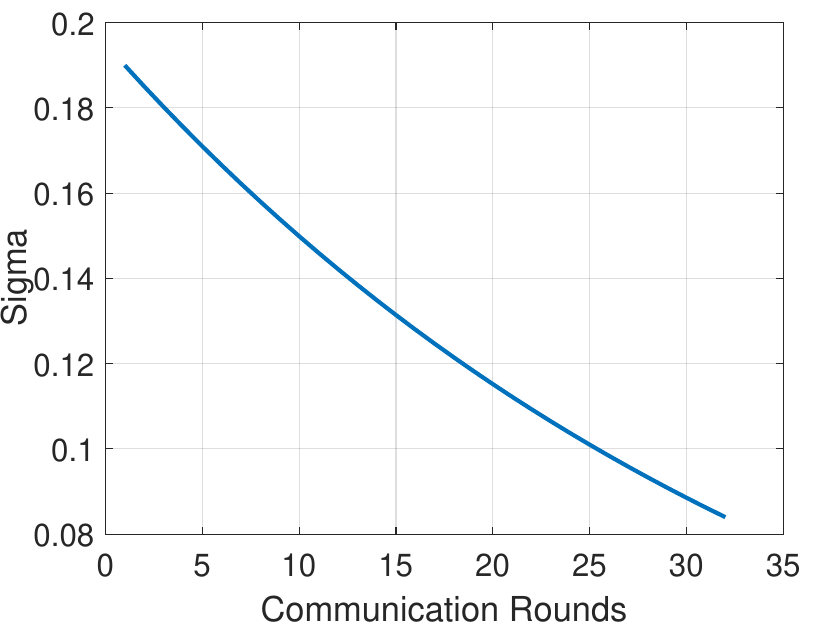}
	}
	\subfigure[CIFAR-10, $(50, 1 \times 10^{-5})$-CLDP.]{
		\includegraphics[width=0.47\columnwidth]{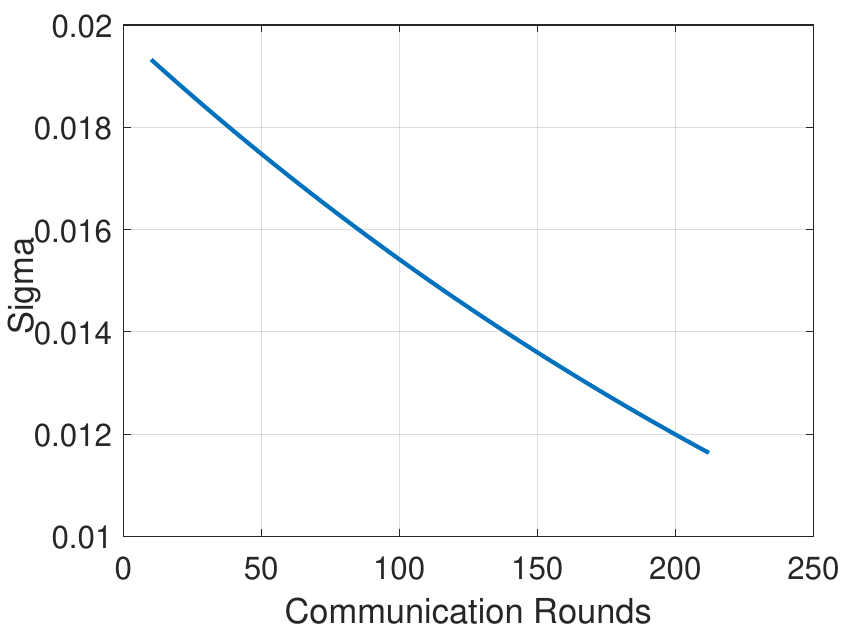}
	}
	\caption{Noise level with rounds of Dynamic Gau-LRQ-SGD.}
	\label{fig:Noise scalar}
\end{figure}

\textbf{Dynamically Determine the Gau-LRQ Parameter $\sigma_k$.} Figure~\ref{fig:Noise scalar} shows the how $\sigma_k$ changes for $k=0,...,K-1$. As we stated before, a small privacy budget leads to a large $\sigma_k$, that is, a large quantization error (i.e., noise) added to the clipped model.
We can see that Dynamic Gau-LRQ-SGD allocates less privacy budget (i.e., add more noise on clipped model) at the early stage of training and reduces noise level as the training goes on. The main reason is that the model update noise in the later stage of training has a greater impact on the convergence error. We have to reduce the variance of model update noise to ensure better convergence of the algorithm. This result is similar to some heuristic work~\cite{gong2020privacy,lee2018concentrated}. Specifically, $\sigma_k$ decreased from 0.19 to 0.08 on MNIST and decreased from 0.020 to 0.012 on CIFAR-10. We can see that the added noise of CIFAR-10 is smaller than that of MNIST. The reason is that the model of CIFAR-10 is more complicated than that of MNIST.

\begin{table}[htbp]
\small
\centering
	\caption{Test Accuracy of Gau-LRQ-SGD on MNIST with Different Clients.}
	\begin{tabular}{ccccc}
		\hline
		Number of Clients& 960 & 1920 & 3840  & 5760\\ 
		\hline
		 Test Accuracy& 0.8711 & 0.9488 & 0.9663  & 0.9745\\ 
		\hline
	\end{tabular}
	\label{tab:differ workers}
\end{table}

\begin{table*}[ht]
\centering
	\caption{Performance Comparison with SOTA on MNIST, CIFAR-10 and CIFAR-100. $-$ means that the algorithm cannot converge. CC denotes the arising communication costs}
   \setlength{\tabcolsep}{1mm}{
	\begin{tabular}{c|cc|cc|cc}
        \hline
        \multirow{2}{*}{Algorithm} & \multicolumn{2}{c|}{MNIST $(3, 1 \times 10^{-5})$}& \multicolumn{2}{c|}{CIFAR-10 $(60, 1 \times 10^{-5})$ }& \multicolumn{2}{c}{CIFAR-100 $(70, 1 \times 10^{-5})$ }\\
		\cline{2-7}
        & CC & Acc & CC & Acc & CC & Acc\\   
		\hline
		Local SGD   & 768 MB  &0.9830 & 1280 GB  &0.9215 & 2400 GB  &0.9207\\
		\hline
        Gau-SGD   & 576 MB  &0.9642& 678.4 GB  &0.7987& 940.8 GB  &0.7691\\
		\hline
		QG-SGD & 36 MB&  0.9579  & 127.2 GB&  0.7879& 235.2 GB&  0.7417\\ 
		\hline
        BQ-SGD & 36 MB&  0.4513 & 127.2 GB&  -& 235.2 GB&  -\\ 
		\hline
	Gau-LRQ-SGD (ours)&    36 MB& \textbf{0.9642} &127.2 GB&  \textbf{0.7987}&235.2 GB&  \textbf{0.7691} \\ 
		\hline
        Dynamic Gau-LRQ-SGD (ours)&  36 MB&  \textbf{0.9711}&  127.2 GB&  \textbf{0.8199}& 235.2 GB&  \textbf{0.7811}\\ 
		\hline
	\end{tabular}}
	\label{tab:Performance comparison with SOTA on datasets}
\end{table*}

\textbf{The Number of Clients}. We take the privacy budget as $(2, 1 \times 10^{-5})$ and fix the number of participated clients each round as $B=80$. Then we vary the total number of clients $N$ from 960 to 5760 and show the test accuracy on MNIST dataset in Table~\ref{tab:differ workers}. From Table~\ref{tab:differ workers}, we can see that more clients can improve the global model's performance. Specifically, when we set the number of clients as 5760, the test accuracy of Gau-LRQ-SGD can achieve 0.9745, almost reaching the accuracy of Local SGD. This suggests that for scenarios where many parties are involved, differential privacy comes at nearly no cost in model performance.

\textbf{Performance Comparison with SOTA.} In Table~\ref{tab:Performance comparison with SOTA on datasets}, we compare the performance of our proposed Gau-LRQ-SGD and Dynamic Gau-LRQ-SGD with some selected algorithms on MNIST, CIFAR-10, and CIFAR-100. The Local SGD, without privacy and communication constraints, provides a benchmark of the testing performance. For MNIST, we set the same privacy budget $(3, 1 \times 10^{-5})$ for all privacy-preserving algorithms. We first observe that our proposed Gau-LRQ-SGD achieves comparable accuracy of 0.9642 as Gau-SGD while the communication cost consumed by Gau-LRQ-SGD is only 36MB, far less than the 576MB of Gau-SGD. 
Dynamic Gau-LRQ-SGD achieves the highest testing accuracy of 0.9711. Compared to Local SGD, we reduced the cost of communication by 95.3\% but only incur a performance impairment of 0.0119. 
In addition, benefit from the strategy of dynamic privacy budget allocation, the performance of Dynamic Gau-LRQ-SGD is higher than that of Gau-LRQ-SGD. For CIFAR-10 and CIFAR-100, we set the privacy budget $(60, 1 \times 10^{-5})$ and $(70, 1 \times 10^{-5})$ for all privacy-preserving algorithms. Similar to the performance on the MNIST dataset, the Gau-LRQ-SGD algorithm outperforms BQ-SGD, GQ-SGD but slightly worse than Dynamic Gau-LRQ-SGD.

%% file: Conclusion.tex
\section{Conclusion}
In this paper, we proposed a new quantization-aided distributed SGD algorithm called Gaussian Layered Randomized Quantization Local SGD (Gau-LRQ-SGD) that could simultaneously achieve communication efficiency and privacy protection. We theoretically capture the trade-offs between communication, privacy, and convergence error. To further enhance the convergence performance, we designed a dynamic privacy budget/quantization step allocation strategy by formulating and solving an optimal problem with respect to minimizing the convergence error bound. Experimental evaluation of various machine learning tasks demonstrates our proposed algorithms outperform the benchmarks.

%% file: Appendix.tex
\section{Appendix}



\subsection{Proof of Lemma 7} \label{proof:lemma7}

We have the following lemma to characterize the aggregated
model update $\bar {\bm \Delta}_k \triangleq \cfrac{1}{B}\sum_{i \in \mathcal{B}_k} \hat{\bm \Delta}_k^{(i)}$ at server:

\begin{lemma}[Unbiasness and Bounded Variance of the aggregated
model update] \label{lemma:aggregated
model update}
	The aggregated model update $\bar {\bm \Delta}_k$ satisfies: 
	\begin{equation}
	\mathbb{E}_{B_i,\mathcal{Q}}[\bar {\bm \Delta}_k] = - \eta \nabla F(\bm{\theta}_k) - \eta\sum_{q=1}^{Q-1} \mathbb{E}_i\nabla F^{(i)}_{k,q},
		\label{eq:unbiassness of update message}
	\end{equation}
	and
	\begin{equation}
	\begin{split}
	 \mathbb{E}_{B_i,\mathcal{Q}}[\| \bar {\bm \Delta}_k \|^2] &\le  \cfrac{Q^2\eta^2\alpha^2}{B} + Q\eta^2 \|\nabla F(\bm{\theta}_k)\|^2 \\
  &+ Q\eta^2\sum_{q=1}^{Q-1}\|\mathbb{E}_i\nabla F^{(i)}_{k,q}\|^2+ \cfrac{d\sigma^2}{B} 
	\end{split}
	\label{eq: Bounded Variance of update message}
	\end{equation}
	where $\mathbb{E}_i\nabla F^{(i)}_{k,q} = \cfrac{1}{B}\sum_{i \in \mathcal{B}_k} \nabla F(\bm{\theta}^{(i)}_{k,q})$.
\end{lemma}

\textbf{Proof:} We have ${\bm \Delta}^{(i)}_k = \bm{\theta}^{(i)}_{k,Q} - \bm{\theta}^{(i)}_{k,0} = -\sum_{q=0}^{Q-1} \eta \bm{g}^{(i)}_{k,q}$, and $\hat{\bm \Delta}_t^{(i)} = \bm \Delta_t^{(i)} + \bm{n}^{(i)}$, where $\bm{n}^{(i)} \sim N(0, \sigma^2\bm{I})$. Hence,

\begin{align*}
	\mathbb{E}_{B_i,\mathcal{Q}}[\bar {\bm \Delta}_k] &= \cfrac{1}{B}\sum_{i \in \mathcal{B}_k} \mathbb{E} \Big[ -\sum_{q=0}^{Q-1} \eta \bm{g}^{(i)}_{k,q}  + \bm{n}^{(i)} \Big]\\
	&\overset{(a)}{=} -\eta \nabla F(\bm{\theta}_k) - \cfrac{\eta}{B} \sum_{i \in \mathcal{B}_k} \sum_{q=1}^{Q-1} \nabla F(\bm{\theta}^{(i)}_{k,q})\\
 & = -\eta \nabla F(\bm{\theta}_k) - \eta \sum_{q=0}^{Q-1} \mathbb{E}_i\nabla F^{(i)}_{k,q}
\end{align*}
where $(a)$ uses Assumption~\ref{ass:stochastic_gradient}. And,
\begin{align*}
	&~~~\mathbb{E}_{B_i,\mathcal{Q}}[\|\bar {\bm \Delta}_k\|^2]\\
	&= \mathbb{E}_{B_i,\mathcal{Q}}[\|\sum_{i \in \mathcal{B}_k} \cfrac{1}{B} [\sum_{q=0}^{Q-1} \eta \bm{g}^{(i)}_{k,q}  + \bm{n}^{(i)}]\|^2]\\ 
 &=  \mathbb{E}_{B_i,\mathcal{Q}}\|\sum_{i \in \mathcal{B}_k} \cfrac{1}{B} \sum_{q=0}^{Q-1} \eta \bm{g}^{(i)}_{k,q}\|^2  + \mathbb{E}_{B_i,\mathcal{Q}} \|\sum_{i \in \mathcal{B}_k} \cfrac{1}{B} \bm{n}^{(i)}\|^2\\
	& \overset{(a)}{\le} Q\eta^2 \sum_{q=0}^{Q-1} \mathbb{E}_{B_i,\mathcal{Q}}\|\sum_{i \in \mathcal{B}_k} \cfrac{1}{B} \bm{g}^{(i)}_{k,q}\|^2  + \cfrac{d\sigma^2}{B} \\
	&\le \cfrac{Q^2\eta^2\alpha^2}{B} + Q\eta^2 \|\nabla F(\bm{\theta}_k)\|^2 + Q\eta^2\sum_{q=1}^{Q-1}\|\mathbb{E}_i\nabla F^{(i)}_{k,q}\|^2+ \cfrac{d\sigma^2}{B} 
\end{align*}
where $(a)$ is the Cauchy–Schwarz inequality.

\subsection{Proof of Theorem 1 and Lemma 6}
\label{proof:The1}
Firstly, we consider function $F$ is $\nu\text{-smooth}$, and use Eq.~ \eqref{eq:smooth_1}, we have:
\begin{equation}\nonumber
\begin{split}
F(\bm{\theta}_{k+1}) &\le F(\bm{\theta}_k) + \nabla F(\bm{\theta}_k)^\text{T} (\bm{\theta}_{k+1}-\bm{\theta}_k)\\
& + \cfrac{\nu}{2} \|\bm{\theta}_{k+1}-\bm{\theta}_k\|^2.
\end{split}
\end{equation}

For the LRQ-SGD, $ \bar {\bm \Delta}_k = \bm{\theta}_{k+1}-\bm{\theta}_k$, so:
\begin{align*}
F(\bm{\theta}_{k+1}) &\le F(\bm{\theta}_k) + \nabla F(\bm{\theta}_k)^\top \bm{\bar \Delta}_k+ \cfrac{\nu}{2} \|\bm{\bar \Delta}_k\|^2.
\end{align*}

Taking total expectations, and using Lemma~\ref{lemma:aggregated
model update}, this yields:

\begin{align}
&\mathbb{E}F(\bm{\theta}_{k+1})-F(\bm{\theta}_k)\nonumber\\
&\le  -\eta \nabla F(\bm{\theta}_k)^\top \Big[\nabla F(\bm{\theta}_k) + \sum_{q=1}^{Q-1} \mathbb{E}_i\nabla F^{(i)}_{k,q}\Big] + \cfrac{\nu}{2} \mathbb{E}\|{\bm \Delta}_k\|^2\nonumber\\
& =-\eta\|\nabla F(\bm{\theta}_k)\|^2 -\eta\sum_{q=1}^{Q-1} \nabla F(\bm{\theta}_k)^\text{T} \mathbb{E}_i\nabla F^{(i)}_{k,q}  + \cfrac{\nu}{2} \mathbb{E}\|{\bm \Delta}_k\|^2 \nonumber\\
& = -\eta\|\nabla F(\bm{\theta}_k)\|^2 - \cfrac{\eta}{2}\sum_{q=1}^{Q-1} \Big[\|\nabla F(\bm{\theta}_k)\|^2 + \|\mathbb{E}_i\nabla F^{(i)}_{k,q}\|^2\nonumber\\
&~~~- \|\nabla F(\bm{\theta}_k)-\mathbb{E}_i\nabla F^{(i)}_{k,q}\|^2 \Big]  + \cfrac{\nu}{2} \mathbb{E}\|{\bm \Delta}_k\|^2\nonumber\\
&\le -\cfrac{\eta(Q+1)}{2}\|\nabla F(\bm{\theta}_k)\|^2 - \cfrac{\eta}{2}\sum_{q=1}^{Q-1}\|\mathbb{E}_i\nabla F^{(i)}_{k,q}\|^2\nonumber\\
&~~~+\cfrac{\eta\nu^2}{2} \sum_{q=1}^{Q-1} \|\bm{\theta}_k-\mathbb{E}_i\bm{\theta}^{(i)}_{k,q}\|^2  + \cfrac{\nu}{2} \mathbb{E}\|{\bm \Delta}_k\|^2
\label{eq:36}
\end{align}

For the third term of above inequality, we have
\begin{align*}
     &~~\sum_{q=1}^{Q-1}\|\bm{\theta}_k-\mathbb{E}_i\bm{\theta}^{(i)}_{k,q}\|^2\\
     &= \eta^2 \sum_{q=1}^{Q-1}\|\sum_{t=0}^{q-1}\mathbb{E}_i \bm{g}^{(i)}_{k,t}\|^2\\
     &\overset{(a)}{\le} \eta^2 \sum_{q=1}^{Q-1}q\sum_{t=0}^{q-1}\|\mathbb{E}_i \bm{g}^{(i)}_{k,t}\|^2\\
     &\le \eta^2 \sum_{q=1}^{Q-1}[q^2\cfrac{\alpha^2}{B}+q\sum_{t=0}^{q-1}\|\mathbb{E}_i\nabla F^{(i)}_{k,t}\|^2]\\
     &\le \eta^2\alpha^2\cfrac{Q(2Q-1)(Q-1)}{6B} + \eta^2 \cfrac{Q(Q-1)}{2}\|\nabla F(\bm{\theta}_k)\|^2\\
     &~~~+ \eta^2 \cfrac{Q(Q-1)}{2}\sum_{q=1}^{Q-1}\|\mathbb{E}_i\nabla F^{(i)}_{k,q}\|^2 
\end{align*}

where $(a)$ is the Cauchy–Schwarz inequality. We plug the above inequality back into Eq.~\eqref{eq:36} and combine Eq.~\eqref{eq: Bounded Variance of update message}, then we can get:

\begin{align*}
    &\mathbb{E}F(\bm{\theta}_{k+1}) - F(\bm{\theta}_k)\\
    &\le -\cfrac{\eta}{2}\Big[Q + 1-\cfrac{\nu^2\eta^2Q(Q-1)}{2} - \nu\eta Q\Big]\|\nabla F(\bm{\theta}_k)\|^2\\
    &-\cfrac{\eta}{2} \Big[ 1-\cfrac{\nu^2\eta^2Q(Q-1)}{2} - \nu\eta Q\Big]\sum_{q=1}^{Q-1}\|\mathbb{E}_i\nabla F^{(i)}_{k,q}\|^2\\
    &+ \cfrac{\nu \eta^2 Q^2\alpha^2}{2B}\Big[\cfrac{\nu \eta(2Q-1)(Q-1)}{6Q}+1\Big]+ \cfrac{\nu d \sigma^2}{2B}
\end{align*}

If the learning rate satisfied $1 \ge \cfrac{\nu^2\eta^2Q(Q-1)}{2} + \nu\eta Q$~\cite{zhou2018convergence} and $\eta \le \cfrac{1}{\nu}$, we have

\begin{equation}\nonumber
\begin{split}
&\mathbb{E}F(\bm{\theta}_{k+1}) - F(\bm{\theta}_k) \\
&\le -\cfrac{\eta Q}{2}\|\nabla F(\bm{\theta}_k)\|^2 + \cfrac{\eta Q^2\alpha^2}{2B}\Big[\cfrac{(2Q-1)(Q-1)}{6Q}+1\Big]+ \cfrac{d\sigma_k^2}{2B\eta}\\
&\overset{(a)}{\le} -\cfrac{\eta Q}{2}\tau^{-k}\|\nabla F(\bm{\theta}_k)\|^2 + \cfrac{\eta Q^2\alpha^2\tau^{-k}}{2B}\Big[\cfrac{(2Q-1)(Q-1)}{6Q}+1\Big]\\
&~~~~~~+ \cfrac{d\tau^{-k}\sigma_k^2}{2B\eta}
\end{split}
\end{equation}
where $(a)$ assumes $-\cfrac{\eta Q}{2}\|\nabla F(\bm{\theta}_k)\|^2 + \cfrac{\eta Q^2\alpha^2}{2B}\Big[\cfrac{(2Q-1)(Q-1)}{6Q}+1\Big]+ \cfrac{d\sigma_k^2}{2B\eta} \le 0$. Applying it recursively, this yields:
\begin{align*}
&\mathbb{E}[F(\bm{\theta}_K)-F(\bm{\theta}_0)] \\
&\le -\cfrac{\eta Q}{2}\sum_{k=0}^{K-1}\tau^{-k}\|\nabla F(\bm{\theta}_k)\|^2
+ \cfrac{d\sum_{k=0}^{K-1}\tau^{-k}\sigma_k^2}{2B\eta}\\
&~~~~~~+ \cfrac{\eta Q^2\alpha^2\sum_{k=0}^{K-1}\tau^{-k}}{2B}\Big[\cfrac{(2Q-1)(Q-1)}{6Q}+1\Big]
\end{align*}

Considering that $F(\bm{\theta}_K) \ge F(\bm{\theta}^*)$, so:
\begin{align}\label{eq:proof_lemma5}
&\frac{1}{\sum_{k=0}^{K-1}\tau^{-k}}\sum_{k=0}^{K-1} \tau^{-k} \|\nabla F(\bm{\theta}_k)\|^2 \nonumber\\
&\le \cfrac{2[F(\bm{\theta}_0)-F(\bm{\theta}^*)]}{Q\eta\sum_{k=0}^{K-1}\tau^{-k}}+ \cfrac{Q\alpha^2}{B}\Big[\cfrac{(2Q-1)(Q-1)}{6Q}+1\Big] \nonumber\\
&~~ +\cfrac{d}{B\eta^2Q\sum_{k=0}^{K-1}\tau^{-k}}\sum_{k=0}^{K-1}\tau^{-k}\sigma_k^2
\end{align}

We complete the proof of Lemma 5. 

$\bullet$ Taking $\sigma_k = \frac{2S_2\sqrt{KB\ln[1/\delta]}}{N\epsilon}$, we conclude the proof of Theorem 1.

$\bullet$ Taking $\sigma_k^2 = \frac{4S_2^2B\ln[1/\delta]}{N^2\epsilon^2} (\sum_{i=0}^{K-1}\tau^{-i/2})\tau^{k/2}$, then the third term of Eq.~\eqref{eq:proof_lemma5} is
\begin{align*}
    &~~~~~\cfrac{d}{B\eta^2Q\sum_{k=0}^{K-1}\tau^{-k}}\sum_{k=0}^{K-1}\tau^{-k}\sigma_k^2\\
    &= \cfrac{d}{B\eta^2Q\sum_{k=0}^{K-1}\tau^{-k}}\sum_{k=0}^{K-1}\tau^{-k}\frac{4S_2^2B\ln[1/\delta]}{N^2\epsilon^2} (\sum_{i=0}^{K-1}\tau^{-i/2})\tau^{k/2}\\
    &= \cfrac{4dS_2^2K\ln[1/\delta]}{\eta^2QN^2\epsilon^2}\cdot\cfrac{AM_K^2(\tau^{-k/2})}{QM_K^2(\tau^{-k/2})}
\end{align*}
where $AM_K(\tau^{-k/2}) =\frac{1}{K} \sum_{k=0}^{K-1} \tau^{-k/2}$ is the Arithmetic Mean, and $QM_K(\tau^{-k/2}) =\sqrt{\frac{1}{K} \sum_{k=0}^{K-1} \tau^{-k}}$ is the Quadratic Mean. We conclude the proof of Theorem 2.

\subsection{Proof of Lemma 4}
\label{proof:lemma5}
We define the privacy loss at $O$ as
\begin{align*}
    c(O,D,D') \triangleq \ln{\cfrac{Pr(\mathcal{M}(D)=O)}{Pr(\mathcal{M}(D')=O)}}
\end{align*}
where $\mathcal{M}$ consists of a sequence of adaptive mechanisms $\{\mathcal{M}_0, \mathcal{M}_1,...,\mathcal{M}_{K-1}\}$. Then the moment generating function evaluated at the value $\lambda$:
\begin{align*}
    \beta_{\mathcal{M}}(\lambda) \triangleq \max_{D,D'} \ln{\mathbb{E}_{O \sim \mathcal{M}(D)}[\exp{(\lambda c(O,D,D'))}]}
\end{align*}
Using the tail bound of $\beta_{\mathcal{M}}(\lambda)$:
\begin{align*}
    \delta &= \min_{\lambda} \exp{(\beta_{\mathcal{M}}(\lambda) - \lambda \epsilon)}\\
    &\overset{(a)}{\le} \min_{\lambda} \exp{(\sum_{k=0}^{K-1}\beta_{\mathcal{M}_k}(\lambda) - \lambda \epsilon)}\\
    &\overset{(b)}{\le} \min_{\lambda}  \exp{(\sum_{k=0}^{K-1} \cfrac{B^2\lambda^2S_{\Delta}^2}{N^2\sigma_k^2} - \lambda \epsilon)}
\end{align*}
where $(a)$ considers the sequence of mechanisms is independent, and $(b)$ using the Lemma 3 in~\cite{abadi2016deep}. $S_{\Delta}=\frac{S_2}{B}$ is the sensitive function of the aggregated model update $\bm{\bar \Delta}_k$. And $\sigma_k^2$ is the variance noise added to $\bm{\bar \Delta}_k$, which is euqally to add $N(0,B\sigma_k^2)$ to each client model updates $\bm{\Delta}^{(i)}_k$. Let $f(\lambda)=\sum_{k=0}^{K-1} \frac{B^2\lambda^2S_{\Delta}^2}{N^2\sigma_k^2} - \lambda \epsilon$, and the minimum value of $f(\lambda)$ is:
\begin{align*}
   \min_{\lambda} f(\lambda) &= -\cfrac{N^2\epsilon^2}{4B^2S_{\Delta}^2\sum_{k=0}^{K-1}\frac{1}{\sigma_k^2}}  = -\cfrac{N^2\epsilon^2}{4S_2^2\sum_{k=0}^{K-1}\frac{1}{\sigma_k^2}}
\end{align*}

Hence,
\begin{align*}
    \delta \le \exp{(-\cfrac{N^2\epsilon^2}{4S_2^2\sum_{k=0}^{K-1}\frac{1}{\sigma_k^2}})}
\end{align*}
That is 
\begin{align*}
    \epsilon = \frac{2S_2\sqrt{\ln[1/\delta]}}{N}\cdot\sqrt{\sum_{k=0}^{K-1}\frac{1}{\sigma_k^2}}
\end{align*}

\subsection{Proof of Lemma 6}\label{proof:lemma6}
Considered that $\hat{\mathbf{\Delta}} = \mathcal{Q}_b[\Tilde{\mathbf{\Delta}} + \bm{n}]$, hence
\begin{align*}
    \mathbb{E}_{\mathcal{Q}} \|\hat{\bm{\Delta}} - \Tilde{\bm{\Delta}}\|^2 & \overset{(a)}{\le} \cfrac{d\mathbb{E}\|\Tilde{\bm{\Delta}} + \bm{n}\|_{\infty}^2}{4l^2} + d\sigma^2\\
    &\le \cfrac{d\|\Tilde{\mathbf{\Delta}}\|_{\infty}^2 + d\sigma^2}{l^2} + d\sigma^2
\end{align*}
where $(a)$ uses the Lemma 1 of~\cite{yan2021dq} and $l = 2^b-1$ is the quantization leval. Taking $b=\log_2{\Big[\frac{\|\Tilde{\mathbf{\Delta}}\|_{\infty}N\epsilon}{2S_2 \sqrt{2\ln{2}KB\ln[1/\delta]}}+1\Big]}$ and $\sigma^2=\frac{4S_2^2KB\ln(1/\delta)}{N^2\epsilon^2}$ into above equation, then
\begin{align*}
    \mathbb{E}_{\mathcal{Q}}\|\hat{\bm{\Delta}} - \Tilde{\bm{\Delta}}\|^2 &\le \frac{4dS_2^2KB\ln(1/\delta)}{N^2\epsilon^2} + \frac{8\ln{2}dS_2^2KB\ln(1/\delta)}{N^2\epsilon^2} \\
    &~~~~~~+ \frac{32d\ln{2}S_2^4K^2B^2\ln(1/\delta)^2}{N^4\epsilon^4\|\Tilde{\mathbf{\Delta}}\|_{\infty}^2}
\end{align*}

Then the analysis of the convergence of QG-SGD is the same as the Proof of Theorem~\ref{proof:The1} by replacing $d\sigma^2_k$ by $\mathbb{E}_{\mathcal{Q}}\|\hat{\bm{\Delta}} - \Tilde{\bm{\Delta}}\|^2$. Then the third term of Eq.~\eqref{eq:proof_lemma5} is
\begin{align*}
    &~~~~~\cfrac{1}{B\eta^2Q} \frac{4dS_2^2KB\ln(1/\delta)}{N^2\epsilon^2} + \cfrac{1}{B\eta^2Q} \frac{2\ln{2}dS_2^2KB\ln(1/\delta)}{N^2\epsilon^2} \\
    &~~~~~~+ \cfrac{1}{B\eta^2Q}\frac{8d\ln{2}S_2^4K^2B^2\ln(1/\delta)^2}{N^4\epsilon^4\|\Tilde{\mathbf{\Delta}}\|_{\infty}^2} \\
    &= \cfrac{4dS_2^2K\ln[1/\delta]}{\eta^2QN^2\epsilon^2} +  \frac{8\ln{2}dS_2^2K\ln(1/\delta)}{\eta^2QN^2\epsilon^2} + \frac{32d\ln{2}S_2^4K^2B\ln(1/\delta)^2}{N^4\epsilon^4\|\Tilde{\mathbf{\Delta}}\|_{\infty}^2\eta^2Q}
\end{align*}

%% file: 0.main.bbl
\begin{thebibliography}{10}
\providecommand{\url}[1]{#1}
\csname url@samestyle\endcsname
\providecommand{\newblock}{\relax}
\providecommand{\bibinfo}[2]{#2}
\providecommand{\BIBentrySTDinterwordspacing}{\spaceskip=0pt\relax}
\providecommand{\BIBentryALTinterwordstretchfactor}{4}
\providecommand{\BIBentryALTinterwordspacing}{\spaceskip=\fontdimen2\font plus
\BIBentryALTinterwordstretchfactor\fontdimen3\font minus \fontdimen4\font\relax}
\providecommand{\BIBforeignlanguage}[2]{{%
\expandafter\ifx\csname l@#1\endcsname\relax
\typeout{** WARNING: IEEEtran.bst: No hyphenation pattern has been}%
\typeout{** loaded for the language `#1'. Using the pattern for}%
\typeout{** the default language instead.}%
\else
\language=\csname l@#1\endcsname
\fi
#2}}
\providecommand{\BIBdecl}{\relax}
\BIBdecl

\bibitem{deng2020edge}
S.~Deng, H.~Zhao, W.~Fang, J.~Yin, S.~Dustdar, and A.~Y. Zomaya, ``Edge intelligence: The confluence of edge computing and artificial intelligence,'' \emph{IEEE Internet of Things Journal}, vol.~7, no.~8, pp. 7457--7469, 2020.

\bibitem{tang2020communication}
Z.~Tang, S.~Shi, X.~Chu, W.~Wang, and B.~Li, ``Communication-efficient distributed deep learning: A comprehensive survey,'' \emph{arXiv preprint arXiv:2003.06307}, 2020.

\bibitem{tao2018esgd}
Z.~Tao and Q.~Li, ``esgd: Communication efficient distributed deep learning on the edge,'' in \emph{USENIX Workshop on Hot Topics in Edge Computing (HotEdge 18)}, 2018.

\bibitem{zhu2020deep}
L.~Zhu and S.~Han, ``Deep leakage from gradients,'' in \emph{Federated learning}.\hskip 1em plus 0.5em minus 0.4em\relax Springer, 2020, pp. 17--31.

\bibitem{fredrikson2015model}
M.~Fredrikson, S.~Jha, and T.~Ristenpart, ``Model inversion attacks that exploit confidence information and basic countermeasures,'' in \emph{Proceedings of the 22nd ACM SIGSAC conference on computer and communications security}, 2015, pp. 1322--1333.

\bibitem{nasr2019comprehensive}
M.~Nasr, R.~Shokri, and A.~Houmansadr, ``Comprehensive privacy analysis of deep learning: Passive and active white-box inference attacks against centralized and federated learning,'' in \emph{2019 IEEE symposium on security and privacy (SP)}.\hskip 1em plus 0.5em minus 0.4em\relax IEEE, 2019, pp. 739--753.

\bibitem{alistarh2017qsgd}
D.~Alistarh, D.~Grubic, J.~Li, R.~Tomioka, and M.~Vojnovic, ``Qsgd: Communication-efficient sgd via gradient quantization and encoding,'' \emph{Advances in Neural Information Processing Systems}, vol.~30, pp. 1709--1720, 2017.

\bibitem{basu2019qsparse}
D.~Basu, D.~Data, C.~Karakus, and S.~Diggavi, ``Qsparse-local-sgd: Distributed sgd with quantization, sparsification, and local computations,'' \emph{arXiv preprint arXiv:1906.02367}, 2019.

\bibitem{nadiradze2021asynchronous}
G.~Nadiradze, A.~Sabour, P.~Davies, S.~Li, and D.~Alistarh, ``Asynchronous decentralized sgd with quantized and local updates,'' \emph{Advances in Neural Information Processing Systems}, vol.~34, pp. 6829--6842, 2021.

\bibitem{dwork2014algorithmic}
C.~Dwork, A.~Roth \emph{et~al.}, ``The algorithmic foundations of differential privacy.'' \emph{Found. Trends Theor. Comput. Sci.}, vol.~9, no. 3-4, pp. 211--407, 2014.

\bibitem{bonawitz2017practical}
K.~Bonawitz, V.~Ivanov, B.~Kreuter, A.~Marcedone, H.~B. McMahan, S.~Patel, D.~Ramage, A.~Segal, and K.~Seth, ``Practical secure aggregation for privacy-preserving machine learning,'' in \emph{proceedings of the 2017 ACM SIGSAC Conference on Computer and Communications Security}, 2017, pp. 1175--1191.

\bibitem{zong2021communication}
H.~Zong, Q.~Wang, X.~Liu, Y.~Li, and Y.~Shao, ``Communication reducing quantization for federated learning with local differential privacy mechanism,'' in \emph{2021 IEEE/CIC International Conference on Communications in China (ICCC)}.\hskip 1em plus 0.5em minus 0.4em\relax IEEE, 2021, pp. 75--80.

\bibitem{mohammadi2021differential}
N.~Mohammadi, J.~Bai, Q.~Fan, Y.~Song, Y.~Yi, and L.~Liu, ``Differential privacy meets federated learning under communication constraints,'' \emph{IEEE Internet of Things Journal}, vol.~9, no.~22, pp. 22\,204--22\,219, 2021.

\bibitem{agarwal2018cpsgd}
N.~Agarwal, A.~T. Suresh, F.~X.~X. Yu, S.~Kumar, and B.~McMahan, ``cpsgd: Communication-efficient and differentially-private distributed sgd,'' \emph{Advances in Neural Information Processing Systems}, vol.~31, 2018.

\bibitem{agarwal2021skellam}
N.~Agarwal, P.~Kairouz, and Z.~Liu, ``The skellam mechanism for differentially private federated learning,'' \emph{Advances in Neural Information Processing Systems}, vol.~34, pp. 5052--5064, 2021.

\bibitem{amiri2021compressive}
S.~Amiri, A.~Belloum, S.~Klous, and L.~Gommans, ``Compressive differentially private federated learning through universal vector quantization,'' in \emph{AAAI Workshop on Privacy-Preserving Artificial Intelligence}, 2021, pp. 2--9.

\bibitem{li2022communication}
Q.~Li, R.~Heusdens, and M.~G. Christensen, ``Communication efficient privacy-preserving distributed optimization using adaptive differential quantization,'' \emph{Signal Processing}, p. 108456, 2022.

\bibitem{yan2023killing}
G.~Yan, T.~Li, K.~Wu, and L.~Song, ``Killing two birds with one stone: Quantization achieves privacy in distributed learning,'' \emph{arXiv preprint arXiv:2304.13545}, 2023.

\bibitem{youn2023randomized}
Y.~Youn, Z.~Hu, J.~Ziani, and J.~Abernethy, ``Randomized quantization is all you need for differential privacy in federated learning,'' \emph{arXiv preprint arXiv:2306.11913}, 2023.

\bibitem{wilson2000layered}
D.~B. Wilson, ``Layered multishift coupling for use in perfect sampling algorithms (with a primer on cftp),'' \emph{Monte Carlo Methods 26 (2000): 141-176.}, vol.~26, pp. 141--176, 2000.

\bibitem{mahmoud2022randomized}
M.~Hegazy and C.~T. Li, ``Randomized quantization with exact error distribution,'' \emph{IEEE Information Theory Workshop (ITW)}, p.~31, 2022.

\bibitem{gong2020privacy}
X.~Y. Gong~M, Feng~J, ``Privacy-enhanced multi-party deep learning.'' \emph{Neural Networks}, vol. 121, pp. 484--496, 2020.

\bibitem{lee2018concentrated}
J.~Lee and D.~Kifer, ``Concentrated differentially private gradient descent with adaptive per-iteration privacy budget,'' in \emph{Proceedings of the 24th ACM SIGKDD International Conference on Knowledge Discovery \& Data Mining}, 2018, pp. 1656--1665.

\bibitem{shi2019distributed}
S.~Shi, Q.~Wang, K.~Zhao, Z.~Tang, Y.~Wang, X.~Huang, and X.~Chu, ``A distributed synchronous sgd algorithm with global top-k sparsification for low bandwidth networks,'' in \emph{2019 IEEE 39th International Conference on Distributed Computing Systems (ICDCS)}.\hskip 1em plus 0.5em minus 0.4em\relax IEEE, 2019, pp. 2238--2247.

\bibitem{rothchild2020fetchsgd}
D.~Rothchild, A.~Panda, E.~Ullah, N.~Ivkin, I.~Stoica, V.~Braverman, J.~Gonzalez, and R.~Arora, ``Fetchsgd: Communication-efficient federated learning with sketching,'' in \emph{International Conference on Machine Learning}.\hskip 1em plus 0.5em minus 0.4em\relax PMLR, 2020, pp. 8253--8265.

\bibitem{mcmahan2017communication}
B.~McMahan, E.~Moore, D.~Ramage, S.~Hampson, and B.~A. y~Arcas, ``Communication-efficient learning of deep networks from decentralized data,'' in \emph{Artificial intelligence and statistics}.\hskip 1em plus 0.5em minus 0.4em\relax PMLR, 2017, pp. 1273--1282.

\bibitem{zhang2020lagc}
J.~Zhang and O.~Simeone, ``Lagc: Lazily aggregated gradient coding for straggler-tolerant and communication-efficient distributed learning,'' \emph{IEEE transactions on neural networks and learning systems}, vol.~32, no.~3, pp. 962--974, 2020.

\bibitem{abadi2016deep}
M.~Abadi, A.~Chu, I.~Goodfellow, H.~B. McMahan, I.~Mironov, K.~Talwar, and L.~Zhang, ``Deep learning with differential privacy,'' in \emph{Proceedings of the 2016 ACM SIGSAC conference on computer and communications security}, 2016, pp. 308--318.

\bibitem{wei2020federated}
K.~Wei, J.~Li, M.~Ding, C.~Ma, H.~H. Yang, F.~Farokhi, S.~Jin, T.~Q. Quek, and H.~V. Poor, ``Federated learning with differential privacy: Algorithms and performance analysis,'' \emph{IEEE Transactions on Information Forensics and Security}, vol.~15, pp. 3454--3469, 2020.

\bibitem{niknam2020federated}
S.~Niknam, H.~S. Dhillon, and J.~H. Reed, ``Federated learning for wireless communications: Motivation, opportunities, and challenges,'' \emph{IEEE Communications Magazine}, vol.~58, no.~6, pp. 46--51, 2020.

\bibitem{kairouz2021distributed}
P.~Kairouz, Z.~Liu, and T.~Steinke, ``The distributed discrete gaussian mechanism for federated learning with secure aggregation,'' in \emph{International Conference on Machine Learning}.\hskip 1em plus 0.5em minus 0.4em\relax PMLR, 2021, pp. 5201--5212.

\bibitem{xiong2016randomized}
S.~Xiong, A.~D. Sarwate, and N.~B. Mandayam, ``Randomized requantization with local differential privacy,'' in \emph{2016 IEEE International Conference on Acoustics, Speech and Signal Processing (ICASSP)}.\hskip 1em plus 0.5em minus 0.4em\relax IEEE, 2016, pp. 2189--2193.

\bibitem{jonkman2018quantisation}
J.~A. Jonkman, T.~Sherson, and R.~Heusdens, ``Quantisation effects in distributed optimisation,'' in \emph{2018 IEEE International Conference on Acoustics, Speech and Signal Processing (ICASSP)}.\hskip 1em plus 0.5em minus 0.4em\relax IEEE, 2018, pp. 3649--3653.

\bibitem{blanchard2017machine}
P.~Blanchard, E.~M. El~Mhamdi, R.~Guerraoui, and J.~Stainer, ``Machine learning with adversaries: Byzantine tolerant gradient descent,'' in \emph{Proceedings of the 31st International Conference on Neural Information Processing Systems}, 2017, pp. 118--128.

\bibitem{yin2018byzantine}
D.~Yin, Y.~Chen, R.~Kannan, and P.~Bartlett, ``Byzantine-robust distributed learning: Towards optimal statistical rates,'' in \emph{International Conference on Machine Learning}.\hskip 1em plus 0.5em minus 0.4em\relax PMLR, 2018, pp. 5650--5659.

\bibitem{zhou2018convergence}
F.~Zhou and G.~Cong, ``On the convergence properties of a k-step averaging stochastic gradient descent algorithm for nonconvex optimization,'' in \emph{Proceedings of the 27th International Joint Conference on Artificial Intelligence}, 2018, pp. 3219--3227.

\bibitem{bottou2018optimization}
L.~Bottou, F.~E. Curtis, and J.~Nocedal, ``Optimization methods for large-scale machine learning,'' \emph{Siam Review}, vol.~60, no.~2, pp. 223--311, 2018.

\bibitem{tang2019doublesqueeze}
H.~Tang, C.~Yu, X.~Lian, T.~Zhang, and J.~Liu, ``Doublesqueeze: Parallel stochastic gradient descent with double-pass error-compensated compression,'' in \emph{International Conference on Machine Learning}.\hskip 1em plus 0.5em minus 0.4em\relax PMLR, 2019, pp. 6155--6165.

\bibitem{yan2022acsgd}
G.~Yan, T.~Li, S.-L. Huang, T.~Lan, and L.~Song, ``Ac-sgd: Adaptively compressed sgd for communication-efficient distributed learning,'' \emph{IEEE Journal on Selected Areas in Communications}, vol.~40, pp. 2678--2693, 2022.

\bibitem{geyer2017differentially}
R.~C. Geyer, T.~Klein, and M.~Nabi, ``Differentially private federated learning: A client level perspective,'' \emph{arXiv preprint arXiv:1712.07557}, 2017.

\bibitem{cheng2022differentially}
A.~Cheng, P.~Wang, X.~S. Zhang, and J.~Cheng., ``Differentially private federated learning with local regularization and sparsification.'' \emph{In Proceedings of the IEEE/CVF Conference on Computer Vision and Pattern Recognition}, pp. 10\,122--10\,131, 2022.

\bibitem{schuchman1964dither}
L.~Schuchman, ``Dither signals and their effect on quantization noise,'' \emph{IEEE Transactions on Communication Technology}, vol.~12, no.~4, pp. 162--165, 1964.

\bibitem{gray1993dithered}
R.~M. Gray and T.~G. Stockham, ``Dithered quantizers,'' \emph{IEEE Transactions on Information Theory}, vol.~39, no.~3, pp. 805--812, 1993.

\bibitem{paladi2021ondemand}
N.~Paladi, M.~Tiloca, P.~N. Bideh, and M.~Hell, ``On-demand key distribution for cloud networks.'' \emph{In 24th Conference on Innovation in Clouds, Internet and Networks and Workshops}, 2021.

\bibitem{zhang2022understand}
X.~Zhang, X.~Chen, M.~Hong, Z.~S. Wu, and J.~Yi, ``Understanding clipping for federated learning: Convergence and client-level differential privacy,'' \emph{International Conference on Machine Learning}, 2022.

\bibitem{yan2021dq}
G.~Yan, S.-L. Huang, T.~Lan, and L.~Song, ``Dq-sgd: Dynamic quantization in sgd for communication-efficient distributed learning,'' in \emph{2021 IEEE 18th International Conference on Mobile Ad Hoc and Smart Systems (MASS)}.\hskip 1em plus 0.5em minus 0.4em\relax IEEE, 2021, pp. 136--144.

\end{thebibliography}
